\documentclass[useAMS,usenatbib]{mn2e}
\usepackage{graphicx}
\usepackage[T1]{fontenc}
\usepackage{aecompl}
\title[A Photometric Variability Survey of M71]{A Search for Photometric Variability Towards M71 with the Near-Infrared Transiting ExoplanetS Telescope}

\author[J. McCormac, I. Skillen, D. Pollacco, F. Faedi, G. Ramsay, V. S. Dhillon, I. Todd, A. Gonzalez]{J. McCormac$^{1,3}$\thanks{E-mail: jmcc@ing.iac.es}, I. Skillen$^{1}$, D. Pollacco$^{2,3}$, F. Faedi$^{2,3}$, G. Ramsay$^{4}$ \and V. S. Dhillon$^{5}$, I. Todd$^{3}$, A. Gonzalez$^{1,6}$\\
$^{1}$Isaac Newton Group of Telescopes, Apartado de Correos 321, E-38700, Santa Cruz de la Palma, Spain\\
$^{2}$Department of Physics, University of Warwick, Coventry CV4 7AL, UK\\
$^{3}$Astrophysics Research Centre, School of Mathematics and  Physics, Queen's University, Belfast, BT7 1NN, UK\\ 
$^{4}$Armagh Observatory, College Hill, Armagh BT61 9DG, UK\\
$^{5}$Department of Physics and Astronomy, University of Sheffield, Sheffield S3 7RH, UK\\
$^{6}$Instituto de Astrof\'isica de Canarias, 38205 La Laguna, Spain }

\begin{document}

\date{Accepted 2013 December 17.  Received 2013 December 13; in original form 2013 August 9}
\pagerange{\pageref{firstpage}--\pageref{lastpage}} \pubyear{2013}

\maketitle

\label{firstpage}

\begin{abstract}
We present the results of a high-cadence photometric survey of an $11\arcmin\times11\arcmin$ field centred on the globular cluster M71, with the Near-Infrared Transiting ExoplanetS Telescope. The aim of our survey is to search for stellar variability and transiting giant exoplanets. This survey differs from previous photometric surveys of M71 in that it is more sensitive to lower amplitude \mbox{($\Delta M\leq0.02$ mag)} and longer period \mbox{($P>2$ d)} variability than previous work on this cluster. We have discovered $17$ new variable stars towards M71 and confirm the nature of $13$ previously known objects, for which the orbital periods of $7$ are refined or newly determined. Given the photometric precision of our high-cadence survey on the horizontal branch of M71, we confirm the cluster is devoid of RR Lyrae variable stars within the area surveyed. We present new $B$ and $V$ band photometry of the stars in our sample from which we estimate spectral types of the variable objects. We also search our survey data for transiting hot Jupiters and present simulations of the expected number of detections. Approximately $1\,000$ stars were observed on the main-sequence of M71 with sufficient photometric accuracy to detect a transiting hot Jupiter, however none were found.
\end{abstract}


\section{INTRODUCTION}		
\label{sec:Introduction}

M71 (NGC 6838) is a sparsely populated, metal-rich Globular Cluster (GC), located at \mbox{R.A. (J2000) 19$^{\mathrm{h}}$53$^{\mathrm{m}}$46\fs49}, \mbox{Dec. (J2000) +18$^{\mathrm{\circ}}$46\arcmin45\farcs1}, at a distance of $3.6$ kpc from the Sun \citep{2000A&AS..144..227G}, in a crowded field not far from the galactic plane, \mbox{$l=56\fdg7$} and \mbox{$b=-4\fdg6$} \citep{2010AJ....140.1830G}. The earliest modern work on variable stars in M71 was carried out by \citet{1953JRASC..47..229S}, who discovered $4$ bright variables in the region towards the cluster, one of which is the long-period semi-regular pulsating giant Z Sagittae and another the semi-detached Algol-type eclipsing binary QU Sge.

The first systematic photometric study of M71 was conducted by  \citet[hereafter AH71]{1971ApJ...167..499A} with the goal of determining the age of the cluster and creating an accurate Colour-Magnitude Diagram (CMD). AH71 carried out $UBV$ photographic and photo-electric observations of M71, calculating the reddening, distance modulus and age of the system to be \mbox{$E(B-V)=0.31\pm0.02$}, \mbox{$(m-M)_{\mathrm{0}}=13.07\pm0.21$} and \mbox{age$=7.6^{+3.1}_{-2.3}$ Gyr}, respectively. \citet{1985AJ.....90...65C} conducted a cluster membership survey of over $350$ stars towards M71 with \mbox{$V<16$} using proper motions. 

In the early 1990s two photometric surveys were carried out in the central region of M71 by \citet{1992AJ....103..460H} and \citet{1994AJ....108.1810Y}, discovering $4$ and $5$ new variable stars, respectively. \citet{1992AJ....103..460H} concluded that one of the $10$ Blue Stragglers (BS) observed is a variable star (H1) of the SX Phoenicis (SX Phe) type. The second variable (H2) is possibly a field dwarf Cepheid ($\delta$ Scuti) and the remaining two variables (H3 \& H4) are possible eclipsing binaries. The goal of the survey by \citet{1994AJ....108.1810Y} was to determine the primordial binary percentage in M71. Three of the variables they discovered are contact binaries (V1, V2 and V5), while the remaining two (V3 and V4) are detached or semi-detached eclipsing binaries. 

\begin{figure}
\rotatebox{270}{\includegraphics[width=63mm]{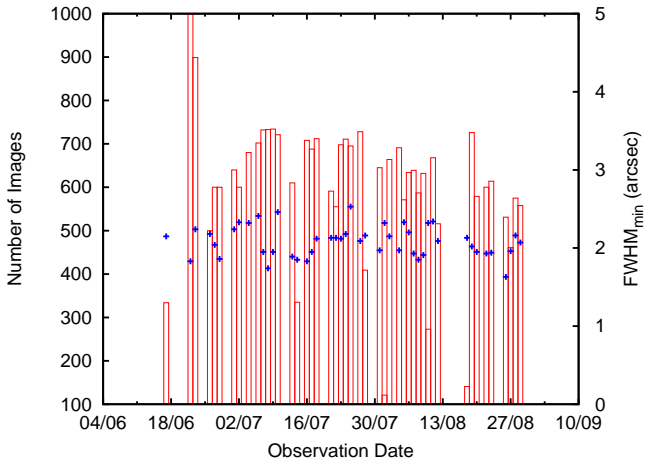}}
\caption{Summary of the NITES survey observations. The red boxes show the number of images acquired per night. Observations on the nights of 2011 Jun 22 and 23, where significantly more images were acquired, were made with $15$ s exposure times. Observations on the other nights were made with $30$ s exposure times. The blue crosses show the minimum FWHM in arc seconds for each night. The minimum FWHM was constrained typically by the lack of fine focus control rather than poor seeing. The focuser on the NITES telescope has since been upgraded.}
\label{figure:observations}
\end{figure}

In the most recent survey for variable stars towards M71, \citet{2000AJ....119.1803P} discovered a further $16$ faint variables with mean magnitudes between \mbox{$15.2<V<20.3$}, $8$ of which are definite variables, $3$ probable and $5$ possible. As previous searches for variable stars in M71 were limited to small areas around the cluster centre, Park \& Nemec primarily conducted their search in the outer regions (out to $r=8.4\arcmin$ from the cluster centre). They identified $4$ possible new SX Phe-type stars (v9, v13, v16 and v22), two of which (v9 and v22) were discovered within \mbox{$1\arcmin$} of the cluster centre. The $16$ faint variables of \citet{2000AJ....119.1803P} combined with the $7$ previously discovered by \citet{1992AJ....103..460H} and \citet{1994AJ....108.1810Y} gave a total of $4$ known bright and $23$ faint variables towards M71 prior to our study.

Whilst the goal of these GC photometric surveys was the detection of variable stars, for more recent surveys the detection of exoplanets has been the prime objective (e.g. 47 Tuc - \citealt{2000ApJ...545L..47G}, \citealt{2005ApJ...620.1043W} and $\omega$ Cen - \citealt{2008ApJ...674.1117W}), although no evidence has been found for an exoplanet orbiting a star in a GC and the hot Jupiter occurrence rate in GCs remains unknown.  

Several transiting exoplanet surveys have also been conducted of younger (age $\leq$ 4 Gyr), more metal-rich \mbox{($-0.23 \leq\mathrm{[Fe/H]}\leq 0.15$)} stars in Open Clusters (OCs), e.g. NGC 6819 \citep{2003MNRAS.340.1287S}, NGC 6940 \citep{2005MNRAS.360..791H}, NGC 7789 \citep{2005MNRAS.359.1096B}, NGC 6791 \citep{2005AJ....129.2856M}, NGC 2158 \citep{2006AJ....131.1090M}, NGC 1245 \citep{2006AJ....132..210B}, Praesepe \citep{2008AJ....135..907P}. None of these surveys found significant evidence of transiting exoplanets. Recently, \citet{2012ApJ...756L..33Q} discovered the first two giant planets orbiting main-sequence stars in the young ($600$ Myr) OC Praesepe using the radial velocity method, indicating that hot Jupiters do form and indeed survive in cluster environments.

\citet{2013MeibomSNGC6811} recently discovered the first two transiting exoplanets orbiting sun-like stars in a cluster environment using data from the \emph{Kepler} spacecraft. The planets, Kepler-66b ($R_{\mathrm{p}}=2.8$ R$_{\oplus}$, $P_{\mathrm{orb}}=17.8$ d) and Kepler-67b ($R_{\mathrm{p}}=2.9$ R$_{\oplus}$, $P_{\mathrm{orb}}=15.7$ d) were discovered in the extremely old ($10^{9}$ yr) OC NGC 6811 as part of the Kepler Cluster Survey \citep{2011AAS...21831103M}. \citet{2013MeibomSNGC6811} find no significant difference between the occurrence rates of Neptune sized planets inside and outside OCs. This strengthens the idea that such planets must be able to survive the high stellar densities and the violent deaths and high energy radiation of massive stars, to which they were inevitably subjected during the early stages of cluster evolution. However, we note that the results of \citet{2013MeibomSNGC6811} are based on low number statistics and that the hot Jupiter occurrence rate in OCs may still differ somewhat from that around field stars. Continued monitoring of many more OCs is required to place more stringent limits on the occurrence rates of exoplanets, of all sizes, in OCs.

Previous surveys in M71 targeted different regions of the cluster with varying levels of photometric accuracy. The survey presented here aims to discover new variable stars and possibly transiting exoplanets in M71. The relatively high metallicity (\mbox{$\mathrm{[Fe/H]}=-0.7$}, \citealt{2011AJ....141...89S}) and lower stellar density of M71 also make it a more favourable host for planetary systems in comparison to other metal-poor, densely packed GCs. M71 also contains more stars than a typical OC, and its apparent diameter \mbox{($\sim7\arcmin$)} is matched well with the field of view (FOV) of the Near-Infrared Transiting ExoplanetS (NITES) telescope (\mbox{$\sim11\arcmin\times11\arcmin$}, see \S \ref{sec:ObservationsDataReduction}) allowing the cluster to be imaged out to $\sim63$ per cent of its tidal radius \mbox{$r_{\mathrm{t}}=8\farcm96$\footnotemark}\footnotetext{http://physwww.mcmaster.ca/$\sim$harris/Databases.html} \citep{1996AJ....112.1487H} in a single pointing. With the NITES telescope we are able to search for lower amplitude, longer period variability than in previous surveys. 

In \S\ref{sec:ObservationsDataReduction} we introduce the NITES telescope, outline our observations of M71 and describe our data reduction processes. In \S\ref{sec:NITESStellarVariablesFinding} we describe our method of detecting stellar variability. In \S\ref{sec:PlanetSearch} we highlight our sensitivity to transiting hot Jupiters in M71 and present simulations of the expected number of detections. In \S\ref{sec:Results} we summarise the results of our survey and  comment on the apparent lack of RR Lyrae variables in M71. In \S\ref{sec:SpectralTypes} we estimate spectral types and discuss cluster membership probabilities for the variable stars in M71. Finally, we close with a summary in \S\ref{sec:Summary}. In Appendix \ref{sec:PhotometricPerformance} we highlight the NITES telescope's photometric performance, comparing it to a theoretical noise model of the system.

\section{THE NITES TELESCOPE, OBSERVATIONS \& DATA REDUCTION}
\label{sec:ObservationsDataReduction}

The NITES telescope is an $f/10$ \mbox{$0.4$ m} Meade LX200GPS Schmidt-Cassegrain telescope. The CCD camera is a Finger Lakes Instrumentation (FLI) Proline 4710 (hereafter PL4710) with a back illuminated \mbox{$1024\times1024$}, \mbox{$13$ $\mu$m} pixel, deep-depleted CCD made by e2v\footnotemark\footnotetext{e2v model 47-10}. The deep depletion CCD suppresses fringing, has a peak quantum efficiency of \mbox{QE $>90$\%} around \mbox{$800$ nm}, and is sensitive out to $1$ micron (where \mbox{QE $\approx25$\%}). The focal length of the telescope is \mbox{$4\,064$ mm} which gives a FOV of \mbox{$11\farcm26\times11\farcm26$}, and a plate scale \mbox{$0.66$ \arcsec~pix$^{-1}$}, respectively. Using a Peltier cooler, PL4710 runs at \mbox{$-40$ \degr C}; at this temperature dark current is noticeable at \mbox{$3.11$ e$^{-}$ pix$^{-1}$ s$^{-1}$}. PL4710 can be read out at pixel rates of $0.75$ or \mbox{$2$ MHz} with resulting gains and read noises of $1.22$ and \mbox{$1.23$ e$^{-}$ ADU$^{-1}$}, and $8.8$ and \mbox{$14.0$ e$^{-}$}, respectively. During our survey of M71 the telescope focus was controlled using a basic system supplied by Meade. The focuser had no absolute position encoder and gave poor focusing repeatability. It has since been upgraded to an FLI Precision Digital Focuser (PDF). The PDF contains an absolute stepper motor, encoded with $7\,000$ steps and a step size of \mbox{$1.25$ $\mu$m}, allowing for more repeatable focus control. The telescope is mounted on a custom wedge made by Maurice Walsh \& Co., Ballynahinch, U.K. The wedge allows for more precise polar alignment over the traditional fixed pier system. The telescope and its control system are housed inside a \mbox{$7$ ft} Astrohaven clamshell dome which is located beside SuperWASP-North \citep{2006PASP..118.1407P} at the Observatorio del Roque de los Muchachos on La Palma. Weather information is taken from the SuperWASP meteorological system in real time. The NITES telescope utilises the DONUTS autoguiding algorithm \citep{2013PASP..125..548M} which calculates guide corrections from the science images directly.

\subsection{Photometry with the NITES Telescope}
\label{subsec:NITESObservations}

\begin{figure}
\rotatebox{270}{\includegraphics[width=62mm]{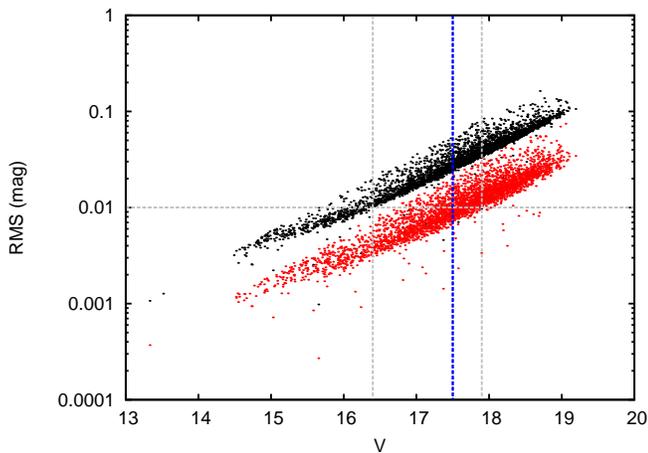}}
\caption{M71 RMS vs $V$ magnitude diagram from a typical night during the NITES survey (\mbox{2011 August 04}). The black points represent the unbinned data while the red points are from data binned into \mbox{$\sim6$ min} bins. The horizontal dashed grey line shows the \mbox{$1$\%} accuracy level required to detect a \mbox{$1$ R$_{\mathrm{Jup}}$} sized planet. The vertical dashed grey lines show the limiting magnitudes at the \mbox{$1$\%} level for the binned and unbinned data. The dashed blue line highlights the magnitude of the main-sequence turn off \mbox{($V=17.5$)} in M71.}
\label{figure:rmsvsmag}
\end{figure}

Remote observations of M71 using the NITES telescope began on 2011 June 17 and continued until 2011 August 29, exposure times of $30$ s were typically used. To increase the sensitivity of the survey, observations were made without a filter. In total the survey collected data over a \mbox{$74$ d} baseline with $47$ nights of observations, equating to \mbox{$228.3$ h} on sky, collecting $28\,340$ images. To monitor the seeing and focus variations during each run and from night to night, the median FWHM of each frame was measured using SExtractor \citep{1996A&AS..117..393B}. A summary of the observations is given in Fig. \ref{figure:observations}. 

Data were reduced using standard routines in IRAF\footnotemark \footnotetext{IRAF is distributed by the National Optical Astronomy Observatories, which are operated by the Association of Universities for Research in Astronomy, Inc., under cooperative agreement with the National Science Foundation.}. The reduced data from the entire survey were then batch processed using the difference-imaging software DIAPL2 (Pych private communication). The software makes use of the method of Optimal Image Subtraction of \citet{1998ApJ...503..325A} and is an improved version of DIAPL \citep{2000AcA....50..421W}. DIAPL2 allows for the creation of smaller image subsections which treat rapidly varying Point Spread Functions (PSFs) better. 

DIAPL2 begins by measuring the average FWHM of all the images to be processed and ranks them in increasing order. The sharpest frames with the lowest background counts are ranked highest and are combined to create the template image of the field to which the others are compared. The highest ranked image is used as the geometric reference to which all other images are aligned. A total of $53$ images met our selection criteria of \mbox{FWHM$<1\farcs98$} ($3$ pixels) and sky background \mbox{$<370$ $\mathrm{e}^{-}$ pixel$^{-1}$}, for template creation. Under-sampled images \mbox{(FWHM$<2$ pixels)} were discarded. The template image was subsequently broken down into $4$ subsections to allow for better PSF fitting and subtraction. Each science image was then split into the same number of subsections and its template subtracted. 

SExtractor was used to extract all the sources in the template image with peak flux per pixel within the linear region of the CCD \mbox{($<45\,000$ e$^{-}$)}. Images taken at elevations \mbox{$<35\degr$} were not used due to dome vignetting and those with focus drifts (median \mbox{FWHM $>4\farcs30$}) were also excluded from the analysis. Approximately \mbox{$8$\%} of the data were lost due to dome vignetting or focus drift. Aperture radii of $10.0$, $8.0$ and $6.0$ pixels ($6\farcs6$, $5\farcs3$ and $4\farcs0$) were tested and a final aperture radius of $6$ pixels ($4\farcs0$) was chosen as it gave the lowest RMS in the reduced light curves. 

To estimate the $V$ band magnitude at peak brightness \mbox{($V_{\mathrm{peak}}$)} of the stars observed with the NITES telescope, $11$ bright photometric standards from Table 1 of AH71 (A1, U, N, A7, A5, Y, X, L, A3, C and A2) were used to transform the white light photometric system to Johnson $V$. As the survey was conducted in white light, no colour information was available making it impossible to convert precisely to Johnson $V$. The method above simply acts as an estimation of the brightness on a magnitude scale.

Mid exposure times were converted to Heliocentric Julian Date (HJD) using the method of \citet{2010PASP..122..935E}. The point-to-point photometric errors from DIAPL2 were systematically larger by a factor of $\sim2$ compared with those from SExtractor. As the errors from SExtractor were closer to those expected from typical CCD noise (see Appendix \ref{sec:PhotometricPerformance}) they were preferred, and the errors from DIAPL2 were scaled using:

\begin{eqnarray}
e_{i,j} &=& \frac{\delta_{i,j}}{\overline{\delta}_{i}} \times E_{i} , \label{eq:real_err} 
\end{eqnarray}

\noindent where $e_{i,j}$ is the corrected error for star $i$ at time $j$, $\delta_{i,j}$ is the error from DIAPL2 for star $i$ at time $j$, $\overline{\delta}_{i}$ is the mean of the errors from DIAPL2 for star $i$ and $E_{i}$ is the error of star $i$ in the reference frame as measured by SExtractor.

\subsection{$BV$ Photometry with the Wide Field Camera}
\label{subsec:WFCObservations}

In order to obtain accurate colour information for all the stars surveyed by the NITES telescope we obtained $6$ $B$ and $6$ $V$ band images of M71 during photometric conditions using the Wide Field Camera (WFC) on the Isaac Newton Telescope on 2013 April 5. The observations in each filter were dithered by $1\arcmin$ steps to cover the gaps between the WFC mosaiced CCDs. The WFC data were reduced using the data reduction package THELI \citep{2005AN....326..432E}, which bias subtracted and flat fielded each of the WFC's $4$ CCDs independently.

As M71 is quite crowded, instrumental magnitudes were obtained using PSF fitting in DAOPHOT \citep{1987PASP...99..191S}. The PSF was modelled using $25$ bright and isolated stars spread over the FOV. Instrumental magnitudes for $27\,185$ stars in the field were obtained and transformed to the standard system using the photometry of GM00. The transformation was carried out using:

\begin{eqnarray}
v_{0} &=& v_{\mathrm{inst}} - \left( X_{v}\times Z_{v} \right), \label{eq:v0}\\ 
b_{0} &=& b_{\mathrm{inst}} - \left( X_{b}\times Z_{b} \right), \label{eq:b0}\\
V-v_{0} &=& a \left( B-V \right) + c_{v}, \label{eq:v} \\
B-V &=& b \left(b_{0} - v_{0} \right) + c_{bv} , \label{eq:bv} 
\end{eqnarray}

\noindent where $v_{\mathrm{inst}}$ and $b_{\mathrm{inst}}$ are instrumental magnitudes, $X_{v}$ and $X_{b}$ are the atmospheric extinction coefficients, $Z_{v}$ and $Z_{b}$ are the average airmass of the mosaic WFC image in each band, $B$ and $V$ are the magnitudes of GM00 on the standard system, $b_{0}$ and $v_{0}$ are the atmospheric extinction-corrected instrumental magnitudes, $a$ and $b$ are the transformation coefficients, $c_{v}$ and $c_{bv}$ are the zero points. As only a small range of airmass was sampled the extinction terms $X_{v}$ and $X_{b}$ were assumed to be $0.2179$ and $0.1036$, respectively. These values were found by interpolating the results from the atmospheric extinction study of La Palma by D. L. King\footnotemark\footnotetext{http://www.ing.iac.es/Astronomy/observing/manuals/ps/\\tech\_notes/tn031.pdf} to the central wavelength of each filter ($B_{\lambda cen}=4298$ \AA~and $V_{\lambda cen}=5425$ \AA).

We calculated $a=0.0015\pm0.0782$, $b=0.83\pm0.02$, $c_{v}=0.58\pm0.10$ and $c_{bv}=0.23\pm0.02$ for the transformations above. After the transformation we find a systematic offset of $+0.128$ mag fainter for the WFC V magnitudes compared to those of GM00. After correcting the systematic offset the magnitudes presented here are found to be in excellent agreement with GM00.

\section{SEARCHING FOR STELLAR VARIABILITY}		
\label{sec:NITESStellarVariablesFinding}

We used a Lomb-Scargle (LS - \citealt{1976Ap&SS..39..447L, 1982ApJ...263..835S,1989ApJ...338..277P}) periodogram to search for stellar variability with periods in the range \mbox{$0.00083<P<100$ d} with frequency steps of \mbox{$\Delta f=0.002$ d$^{-1}$}. For each periodogram, we estimated the power detection threshold corresponding to a false alarm probability (FAP) of 0.1\%, according to the number of independent frequencies $N_{freq}$ used (see \citealt{1999ApJ...526..890C}). We tested our FAP performing extensive simulations on a set of $1\,000$ synthetic light curves free from astrophysical periodic signals, modelled using real NITES light curves (time stamps, data gaps, data length). Because our data contain real periods at the $n$ and $n/x$ periods (mostly caused by our window function) we also used the CLEAN \citep{1987AJ.....93..968R} algorithm to confirm our period detection.

Additionally, light curves with significant \mbox{(FAP $\leq$ 0.1\%)} periodicity from our LS analysis were analysed further using Phase Dispersion Minimisation (PDM - \citealt{1978ApJ...224..953S}). The range of periods searched with PDM were restricted to \mbox{$\pm0.1$\%} of the period detected by the LS and CLEAN analyses. It was found that PDM typically returned a better fitting period, especially in the case of non-sinusoidal variability. Variable stars displaying evidence of multiple periods were also analysed using a Fourier power spectrum in the PERIOD04 program \citep{Period04Lenz}. \citet{Breger1993} and \citet{Kuschnig1997} show that for a significant detection a $SNR>4$ is required for a given peak in the power spectrum, hence we adopt the same criterion in our multifrequency analysis (see \S\ref{subsec:StellarVariablesNew}). A recent comparison of period fitting algorithms by \citet{2013MNRAS.434.3423G} showed that different types of variability in data of differing quality -- both in terms of time sampling and noise characteristics -- may have their periodicity more efficiently recovered by different algorithms. Here we choose to concentrate on periodic variability (stellar pulsations, eclipsing binaries as well as transiting exoplanets) and hence adopt the traditional algorithms, LS and PDM, and Box Least Squares (BLS, see \S \ref{subsec:PlanetSearchTransitDetectionMethod}) in the case of searching for transiting exoplanets. We note however, that these algorithms are not perfect; e.g. the well known half-period aliasing problem of eclipsing binaries when using an LS analysis.

The error on each period was calculated using a bootstrapping technique. A random number generator was used to randomly resample, with replacement, each unfolded light curve $500$ times, resulting in a series of light curves in which several data points may have been selected more than once and others not at all. PDM was run on this series of resampled light curves and the error on the period was taken as the standard deviation in the spread of periods measured in the resampled series. In the case of eclipsing binaries showing ellipsoidal variations and/or secondary eclipses, the strongest PDM period is often half the true orbital period. These light curves were easily identified by eye and their PDM period doubled. 

\section{DETECTABILITY OF TRANSITING HOT JUPITERS IN M71}
\label{sec:PlanetSearch}

To detect the transit of a typical hot Jupiter requires photometric precision at the \mbox{$1$\%} level or better. Figure \ref{figure:rmsvsmag} shows a typical RMS vs magnitude diagram of M71 taken during dark time with the NITES telescope. The unbinned data from the night of 2011 August 04 (black points) are of insufficient accuracy around the main-sequence turn off (\mbox{MSTO, $V=17.5$}) to detect a transiting hot Jupiter. Figure \ref{figure:noise} shows that the NITES telescope is essentially free from correlated noise, hence binning of our data points, as done by the BLS fitting algorithm (see \S\ref{subsec:PlanetSearchTransitDetectionMethod}) to increase our photometric accuracy is justified. The data from \mbox{2011 August 04} were subsequently binned tenfold into \mbox{$\sim6$ min} bins and the RMS vs magnitude diagram was recreated (see Fig. \ref{figure:rmsvsmag} - red points). At this level of binning it is clear that our variability survey is sensitive to transiting hot Jupiters at the top of the MS in M71. Since our survey only samples the bright end of the MS in M71, our targets are those with the least photon noise which allows for easier transit detection.

\subsection{Transiting Exoplanet Expectations}
\label{subsec:PlanetSearchExpectations}

\begin{figure}
\rotatebox{270}{\includegraphics[width=6cm, height=9cm, trim=0mm 10mm 0mm 0mm]{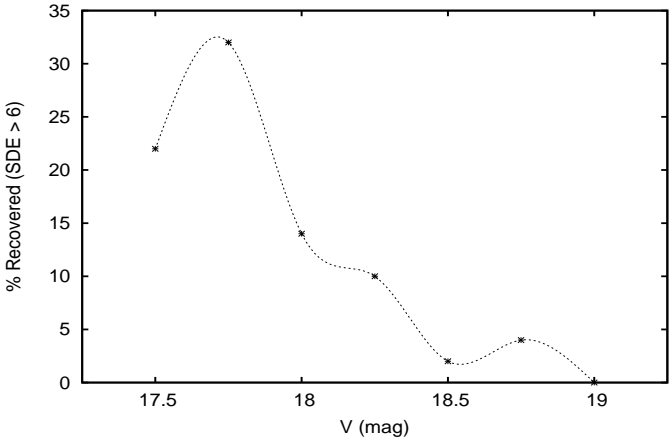}}
\caption{BLS detection efficiency per magnitude bin for a \mbox{$1.3$ $R_{\mathrm{Jup}}$} planet on a \mbox{$P=3.2$ d} orbit. Detections are only assumed real if \mbox{$SDE>6$} (see \S\ref{subsec:PlanetSearchTransitDetectionMethod}).}
\label{figure:bls}
\end{figure}

In this section we describe the process of creating a series of simulated NITES light curves containing hot Jupiter-like transits of differing depths and durations. The goal is to simulate the possible transit signals we might detect in our survey data and determine our detection success rate using the transit detection algorithm described in \S\ref{subsec:PlanetSearchTransitDetectionMethod}. Given the hot Jupiter median radius and period ($1.256$ $R_{\mathrm{Jup}}$ and $3.235$ d, August 2013. See exoplanet encyclopedia\footnotemark\footnotetext{http://www.exoplanet.eu}) we investigated our sensitivity in the case of a hot Jupiter with $R_{\mathrm{p}} = 1.3$ $R_{\mathrm{Jup}}$ and a period of $P= 3.2$ d.

We adopted the estimates for $R_{\mathrm{s}}$ and $M_{\mathrm{s}}$ as in 47 Tuc (see \citealt{2001ApJ...556..322B} and also \citealt{2005ApJ...620.1043W}) for the simulations of exoplanet detectability presented here. Excluding the effects of limb darkening, the transit depth, $\delta M$ and duration, $t_{\mathrm{trans}}$, in hours for the simulated light curves can be estimated using:

\begin{eqnarray}
\delta M &\sim& \left(\frac{R_{\mathrm{p}}}{R_{\mathrm{s}}}\right)^{2}, \label{eq:transdepth2}\\
t_{\mathrm{trans}} &=& 1.412\ M_{\mathrm{s}}^{-1/3}\ R_{\mathrm{s}}\ P^{1/3}, \label{eq:translength2}
\end{eqnarray}

\noindent where $R_{\mathrm{p}}$ and $R_{\mathrm{s}}$ in Eq  \ref{eq:transdepth2} are the radii of the planet and star respectively, and $M_{\mathrm{s}}$ and $R_{\mathrm{s}}$ in Eq \ref{eq:translength2} are the mass and radius of the star in solar units and $P$ is the orbital period in days \citep{2000ApJ...545L..47G}.

For a given $V$ magnitude each light curve was created with depth and duration from Table \ref{table:2}. The synthetic set spans the length of the survey with a time sampling of $0.0005$ d, chosen in order to match the real NITES time sampling. The first mid-transit time for each individual light curve was randomly selected from the three days before our observing window. Gaussian noise was then added to individual light curves where the RMS is the limiting accuracy of the NITES survey in each mag bin (see Table \ref{table:2}).

The synthetic light curves were compared to the real times of observation during the survey and points which did not correspond to time on sky were excluded. As observations were carried out only when conditions were favourable no explicit simulations of weather and observing conditions were performed. The process was repeated $50$ times for each $V$ magnitude bin (see Table \ref{table:2}), resulting in $350$ simulated light curves with randomly occurring transits throughout. As can be seen in Table \ref{table:2}, the depths of any transits detectable by the NITES telescope are expected to range from \mbox{$\sim0.01-0.03$ mag}.

\begin{table}
\caption{Simulated light curve parameters calculated using Eqs \ref{eq:transdepth2} \& \ref{eq:translength2} and the estimates of stellar mass and radius of \citet{2001ApJ...556..322B}. The final column is the number of stars in each magnitude bin of the NITES survey with \mbox{RMS $<$ transit depth for a $1.3$ R$_{\mathrm{Jup}}$} planet.}
\label{table:2}
\begin{tabular}{@{}lcccc}
\hline\hline
V (mag) & Transit  & Transit  & RMS$_{\mathrm{lim}}$ & N$_{\mathrm{stars}}$ \\
 & Depth (mag) & Duration (h) & (mag) &  \\
\hline
17.50	& 0.01074 & 2.79 & 	0.007	& 	210	\\
17.75	& 0.01369 & 2.48 & 	0.009	&	255	\\
18.00	& 0.01635	 & 2.29 & 	0.011	&	268	\\
18.25	& 0.01942	 & 2.12 & 	0.013	&	189	\\
18.50	& 0.02280 & 1.97 & 	0.017	&	115	 \\
18.75	& 0.02617	 & 1.86 & 	0.022	&	36	 \\
19.00	& 0.02958 & 1.76 & 	0.030 	&	1 	 \\
\end{tabular}
\end{table}

The simulated light curves were analysed with the BLS algorithm explained in \S\ref{subsec:PlanetSearchTransitDetectionMethod}. The expected transit depth in each magnitude bin from Table \ref{table:2} was used to separate stars suitable for transit searching from those with insufficient photometric accuracy. The final column in Table \ref{table:2} gives the total number of stars per magnitude bin with photometric precision better than the corresponding depth of a \mbox{$1.3$ R$_{\mathrm{Jup}}$} planet. A total of $1\,074$ stars were observed on the MS of M71 down to \mbox{$V\sim19.0$} with sufficient accuracy to discover a transiting hot Jupiter. 

There are currently conflicting numbers regarding the hot Jupiter occurrence rate around Sun-like stars in the Solar neighbourhood. \citet{2012ApJ...753..160W} summarise the situation and show that the rate differs by up to a factor of $4$ depending on the source. Results from the \emph{Kepler} \citep{2012ApJS..201...15H} and \emph{OGLE-III} \citep{2006AcA....56....1G} transit surveys (adjusted for all orbital geometries, transiting and non-transiting) disagree with those from the radial velocity (RV) surveys \citep{2005PThPS.158...24M,2008PASP..120..531C,2011arXiv1109.2497M,2012ApJ...753..160W} by a factor of $2-3$. \citet{2012ApJ...753..160W} suggest part of this may stem from the likely lower metallicity in the vicinity of the distant \emph{Kepler} field objects. These distant objects will have significant heights above the galactic plane, possibly separating the population from that of the RV surveys, by age and potentially, metallicity. \citet{BeattyGaudi2008} show that in the S/N limited case (i.e. \emph{Kepler}) the transit detection rate goes as $R_{p}^{6}$, hence the properties of the RV and transit survey planets are considerably different, notably the larger radii and shorter orbital periods discovered by the transit surveys.

\begin{figure*}
\rotatebox{270}{\includegraphics[width=10cm,height=18cm,trim=10mm 10mm 10mm 20mm]{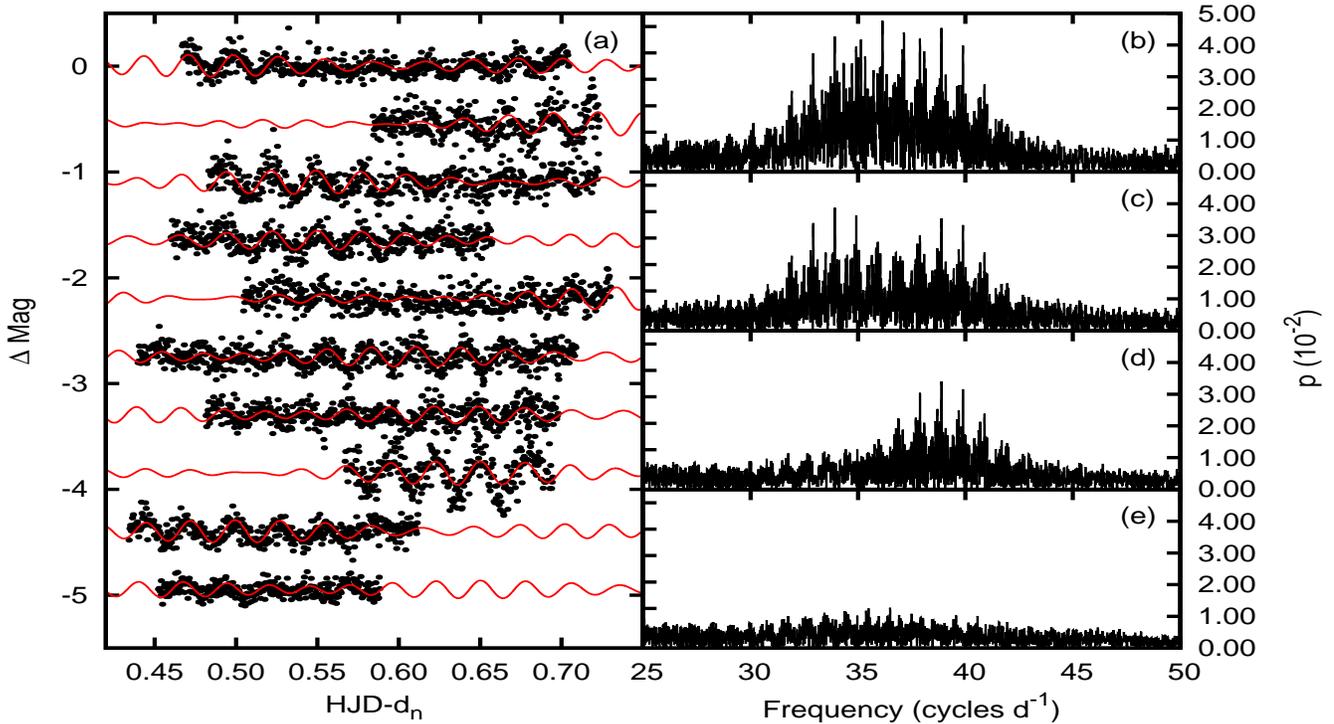}}
\caption{Light curves (left) and power spectra (right) of QU Sge. Three peaks ($f_{1}=33.9167$, $f_{2}=36.1381$ and $f_{3}=38.8864$ cycles d$^{-1}$) are found in the power spectrum of QU Sge out-of-eclipse data. Panel (a): Out-of-eclipse SX Phe pulsations with the best fitting model over plotted. Beating of the frequencies detected is clearly evident on nightly inspection. The data for different nights in Panel (a) have been offset by multiples of $0.55$ mag for clarity, $d_{n}$ is the Julian Day for a given night. Panel (b): power (p) spectrum of the raw data of QU Sge. Panel (c): power spectrum after the strongest peak ($f_{2}=36.1381$ cycles d$^{-1}$) has been removed. Panels (d) and (e): power spectra after the removal of the remaining two significant peaks, $f_{1}=33.9167$ cycles d$^{-1}$ and $f_{3}=38.8864$ cycles d$^{-1}$, respectively.}
\label{figure:qusge_sxphe_combined}
\end{figure*}

If we assume an average of only \emph{Kepler} and \emph{OGLE-III} occurrence rates we would expect there to be $4$ hot Jupiters per thousand MS stars. Assuming \mbox{$10$\%} of those transit, we expect there to be $0.4$ detectable transiting hot Jupiters in the NITES survey. \citet{Johnson2010} suggest that the giant planet occurrence rate is dependent on both the host star mass \emph{and} metallicity. However, Eq 8 of \citet{Johnson2010} does not account for the fraction of giant planets of short period orbits. In reality, the likelihood of detecting a hot Jupiter on the MS of M71 depends on several factors - metallicity, stellar mass and orbital period - the details of which remain uncertain. This makes the search for transiting giant planets in M71 an interesting endeavour.

Using the BLS detection efficiencies shown in Fig. \ref{figure:bls} we calculate an average of a \mbox{$\sim10$\%} chance of recovering a transiting hot Jupiter signal over the MS magnitude range $17.5<V<18.75$. No significant detections were returned from the synthetic light curve analysis at $V=19.0$.

\subsection{Transit Detection Method}			
\label{subsec:PlanetSearchTransitDetectionMethod}

The light curves were searched for transit signals using the BLS algorithm of \citet{2002A&A...391..369K}. We searched for transit durations between $0.01 < t_{frac} < 0.1$ and periods between $0.5 < P< 10$ d in $2\,000$ frequency steps of $0.0001$ d$^{-1}$. The Signal Detection Efficiency (SDE) was calculated for each BLS power spectrum using Eq $6$ of \citet{2002A&A...391..369K}, and we assume the same criterion \mbox{$SDE>6$} to represent a significant detection.

\section{RESULTS \& DISCUSSION}
\label{sec:Results}

Our search for variable stars has discovered $17$ new variables towards M71, and $7$ of the previously proposed variables have been confirmed and had their periods determined or refined. The new variables all lie within the cluster's projected tidal radius \mbox{$r_{\mathrm{t}} = 8\farcm96$}. They have been given numbers ranging from v24 to v40, in keeping with the nomenclature of \citet[hereafter PN00]{2000AJ....119.1803P}. Section \ref{subsec:TransitingHotJupitersInM71} highlights the results from our search for hot Jupiters in the NITES survey data. A brief description of only the remarkable - in the sense that their period or type of variability is found to be different here than that published in the literature - previously discovered variables (QU Sge, S4, v10 and v16) is given in \S\ref{subsec:StellarVariablesPrevious}. The details of the others (v1, v2, v3, v4, v5, v19, v20, v21 and v23) are summarised in Tables \ref{table:3} and \ref{table:4}. Section \ref{subsec:StellarVariablesNew} gives an overview of the new variable stars discovered here (v24-v40). Folded light curves for those with a single periodicity detected are shown in Figs \ref{figure:variables1} \& \ref{figure:variables2}. For clarity the light curves have been averaged into bins of $0.001$ in phase and the error bars have been excluded. Figures \ref{figure:v26-v31} and \ref{figure:v34-v40} show the light curves of the multiply periodic variables v26, v27, v28, v31, v34, v35, v36 and v40. For clarity, in Figs \ref{figure:v26-v31} and \ref{figure:v34-v40} the data points have been averaged into $50$ bins per night and the error bars have also been excluded. A summary of the frequencies detected in the multiply periodic variables is given in Table \ref{table:freqs}. A finding chart for each of the new variables is given in Fig. \ref{figure:finders}. We would like to note that it is notoriously difficult to perform precise photometry is crowded stellar environments and in the presence of saturated stars located close to variable objects (e.g. \citealt{2013A&A...553A.111S}). The inevitable effects of blending may influence to a certain degree the amplitude and/or the period of variable stars in crowded areas or regions surrounding saturated objects.

\subsection{Transiting Hot Jupiters Towards M71}
\label{subsec:TransitingHotJupitersInM71}

A total of $1\,074$ light curves were flagged as having sufficient photometric accuracy to detect a \mbox{$1.3$ R$_{\mathrm{Jup}}$} planet on a \mbox{$3.2$ d} orbit and were analysed using the BLS method above. A total of $5$ objects showed significant detections \mbox{(SDE $>6$)}, two of which are variable stars (v28 and v30 - see \S\ref{subsec:StellarVariablesPrevious}) and one is a probable eclipsing binary of PN00 (v10 - confirmed here in \S\ref{subsec:StellarVariablesNew}). The remaining two when folded on the detected period showed none of the characteristics of a transiting hot Jupiter; shallow, flat bottomed and short duration transits with respect to the period of variability. Although our sample is of more than $1\,000$ stars, in order to place robust statistical constraints on the hot Jupiter frequency in a relatively metal-rich GC, a much larger sample of the cluster's MS is required.

\subsection{Previously Discovered Variables Towards M71}		
\label{subsec:StellarVariablesPrevious}

\begin{figure*}
\includegraphics[width=18cm,height=18cm,trim=0mm 0mm 0mm 0mm]{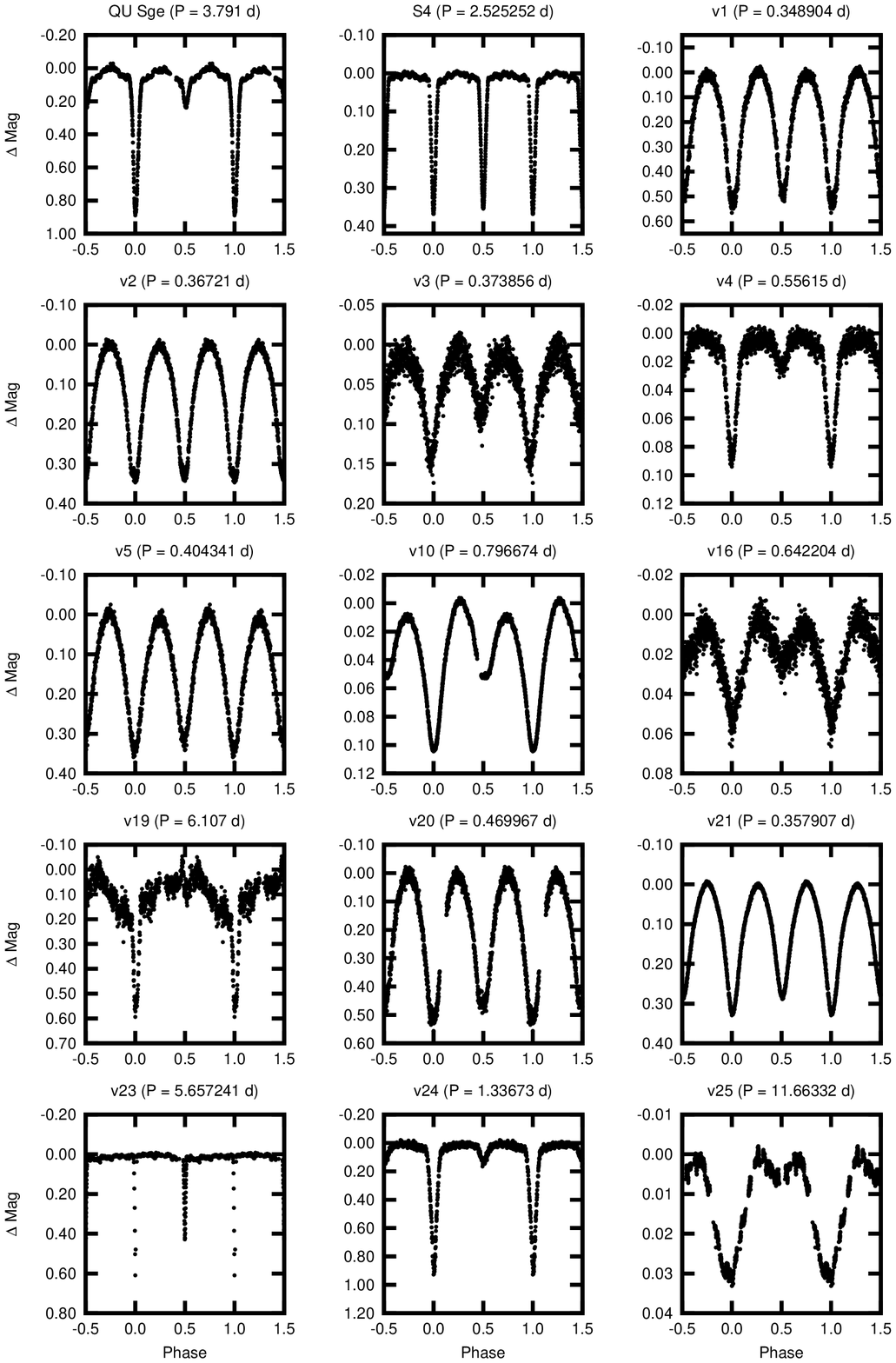}
\caption{Folded light curves for the singly periodic, previously known variable stars QU Sge, S4, v1, v2,  v3, v4, v5, v10, v16, v19, v20, v21 and v23, the newly confirmed Algol-type eclipsing binary v24 and the newly discovered eclipsing binary v25. The periods of S4, v10, v16, v19, v20, v21 and v23 have been improved through the analysis here, v23 is also confirmed as an Algol-type eclipsing binary. PN00 suggested v16 could be an SX Phe-type variable but from the analysis presented here it appears to be eclipsing binary.}
\label{figure:variables1}
\end{figure*}

QU Sge is a semi-detached Algol-type eclipsing binary containing an SX Phe-type component discovered by \citet{2006ApJ...636L.129J}. This was the first pulsating component of an eclipsing binary found in a GC. \citet{2006ApJ...636L.129J} refined the orbital period of QU Sge to \mbox{$P=3.790818 \pm 0.000012$ d}. The eclipsing binary light curve was subtracted from the data revealing short-period \mbox{($P\sim0.03$ d)} low-amplitude \mbox{($\sim0.024$ mag)} variations which are consistent with SX Phe-type variability. \citet{2006ApJ...636L.129J} find two peaks in the power spectrum of QU Sge, $f_{1}=35.883$ cycles d$^{-1}$ and $f_{2}=39.867$ cycles d$^{-1}$. As the ratio of the frequencies detected is larger than $0.8$ (typical value for fundamental and first-overtone modes assumed by \citealt{2006ApJ...636L.129J}), they conclude that at least one of the two periods originates in a non-radial mode. 

A period of \mbox{$P = 3.79100 \pm 0.00061$ d} for the eclipses of QU Sge (see Fig. \ref{figure:variables1}) was found here, which is in agreement with the period \mbox{$P = 3.790818$ d} of \citet{2006ApJ...636L.129J}. Nightly inspection of our QU Sge light curves also showed SX Phe-type variations. Non-eclipsing data were normalised with a second order polynomial and converted to differential magnitudes to highlight the variability. The results are shown in Fig. \ref{figure:qusge_sxphe_combined}. The out-of-eclipse data were analysed collectively using a Fourier analysis. We find a non-equally spaced triplet in the power spectrum of QU Sge, with $f_{1}=33.9167$ cycles d$^{-1}$, $f_{2}=36.1381$ cycles d$^{-1}$ and $f_{3}=38.8864$ cycles d$^{-1}$, two of which are similar to those of \citet{2006ApJ...636L.129J} (see Fig. \ref{figure:qusge_sxphe_combined}). The fundamental period of high-amplitude ($\Delta V\approx0.2-0.3$) $\delta$ Scuti stars is shorter than the first over-tone pulsation period by a factor of $0.775$ \citep{2005A&A...440.1097P}. The frequency ratios, $f_{1}/f_{2}=0.9385$, $f_{1}/f_{3}=0.8722$ and $f_{2}/f_{3}=0.9293$, from our analysis are also larger than $0.775$, hinting that at least two of the detected periods originate from non-radial modes. 

\begin{figure*}
\rotatebox{270}{\includegraphics[width=12cm, height=17cm,trim=0mm 0mm 0mm 00mm]{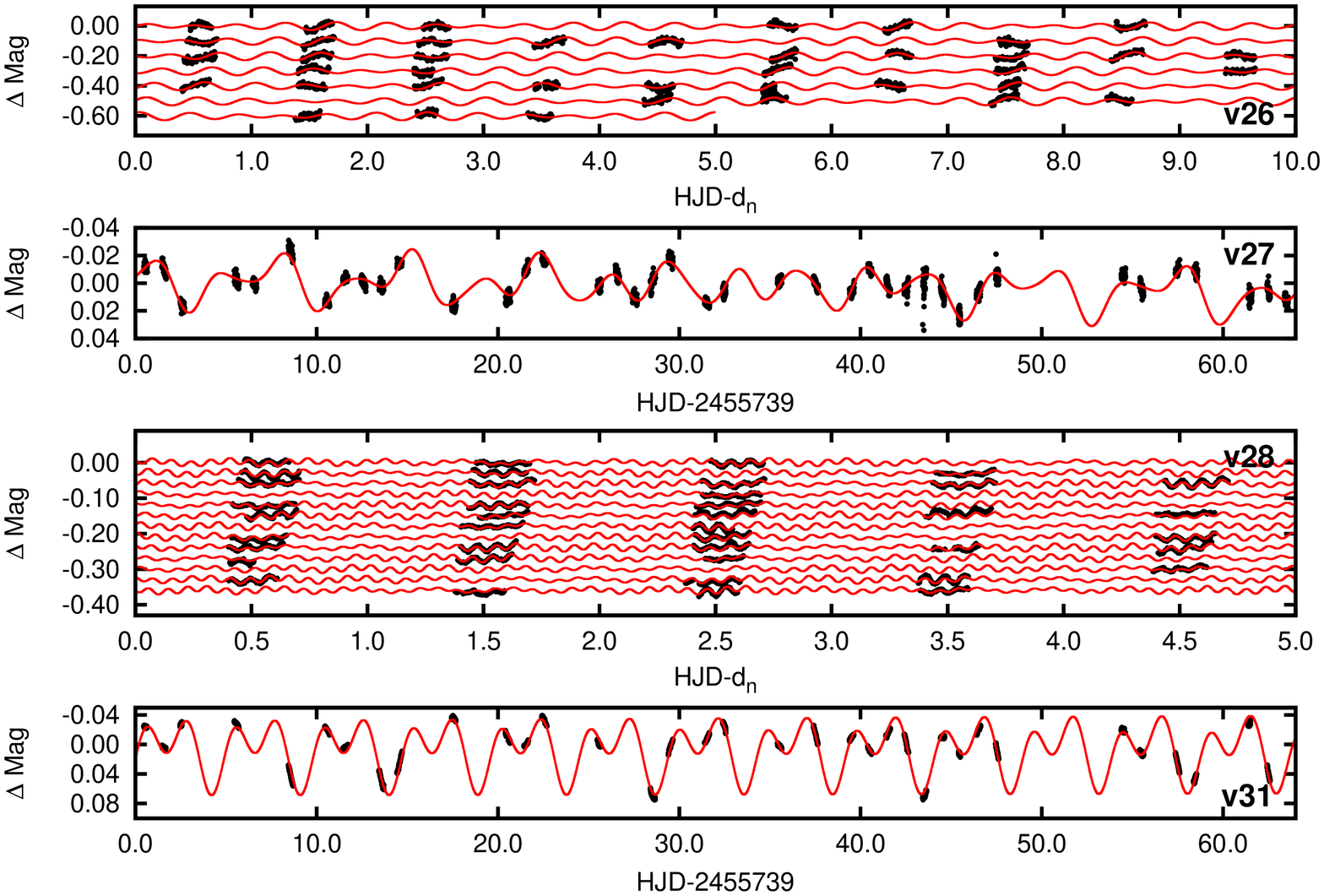}}
\caption{Light curves for the multiply periodic variables v26, v27, v28 and v31, over plotted with the best fitting models from the Fourier analyses. The data have been combined into $50$ binned points per night. For clarity, the data for v26 and v28 are shown in groups of $10$ and $5$ nights (starting at HJD 2455739, where $d_{n}$ is the Julian Day for a given night) and each groups of nights have been offset by integer numbers of $0.10$ and $0.03$ mag, respectively. The significant frequencies detected for each object are given in Table \ref{table:freqs}.}
\label{figure:v26-v31}
\end{figure*}

The amplitude of the variations observed here are much larger, \mbox{$\sim0.2$ mag} compared to \mbox{$24$ mmag} of \citet{2006ApJ...636L.129J} and is seen to vary over time (see Fig. \ref{figure:qusge_sxphe_combined}). The greater amplitude observed here is likely caused by the modulation of the light curve by the three different, but closely separated frequencies. The difference in photometric bands between our observations and those of \citet{2006ApJ...636L.129J} (white light vs $V$) may also contribute to the differences observed in pulsation amplitudes. Assuming we have detected radial pulsations and the fundamental frequency we use the period-luminosity relation of \citet[their Eq 1]{2011AJ....142..110M} to estimate the distance to QU Sge. We find $d=1.68\pm0.07$ kpc, which places QU Sge in the foreground of M71 and agrees with the marginal cluster membership probability ($51$\%) of \citet{1985AJ.....90...65C}. If the first overtone has been detected the distance above will be shorter by $\sim15\%$.

Our NITES observations of several other previously known variables have revealed some interesting results. S4 was originally proposed to be an RR Lyrae variable by \citet{1973PDDO....3....6S}, however it actually appears to be a detached eclipsing binary with a period of \mbox{$P = 2.525252\pm0.000033$ d}, see Fig. \ref{figure:variables1}. v10 was discovered by PN00 as a possible W UMa-type contact binary system with a period of \mbox{$P = 0.76842$ d}. However, the period found here \mbox{($P=0.796674 \pm 0.000033$)} differs significantly from that of PN00 by \mbox{$\sim 41$ min}. PN00 quote no errors on the period of v10. We suspect that the difference most likely stems from their incomplete phase coverage of their observations, see Fig. \ref{figure:variables1}. Finally, the variable star v16 was suggested to be an SX Phe-type variable by PN00. Our NITES observations show v16 to have variations of differing depths, see Fig. \ref{figure:variables1}. v16 therefore appears to be a possible contact binary rather than a SX Phe variable.

\subsection{New Variables Towards M71}			
\label{subsec:StellarVariablesNew}

{\bf{v24}}: The first new eclipsing binary system found in the direction of M71 in over $10$ years. We find a period using PDM of \mbox{$P = 1.33673\pm0.00029$ d}. v24 appears to be a detached eclipsing binary system with a primary eclipse depth approximately twice as large as the secondary (see Fig. \ref{figure:variables1}).

\begin{figure*}
\includegraphics[width=18cm,height=12cm,trim= 0mm 100mm 10mm 0mm]{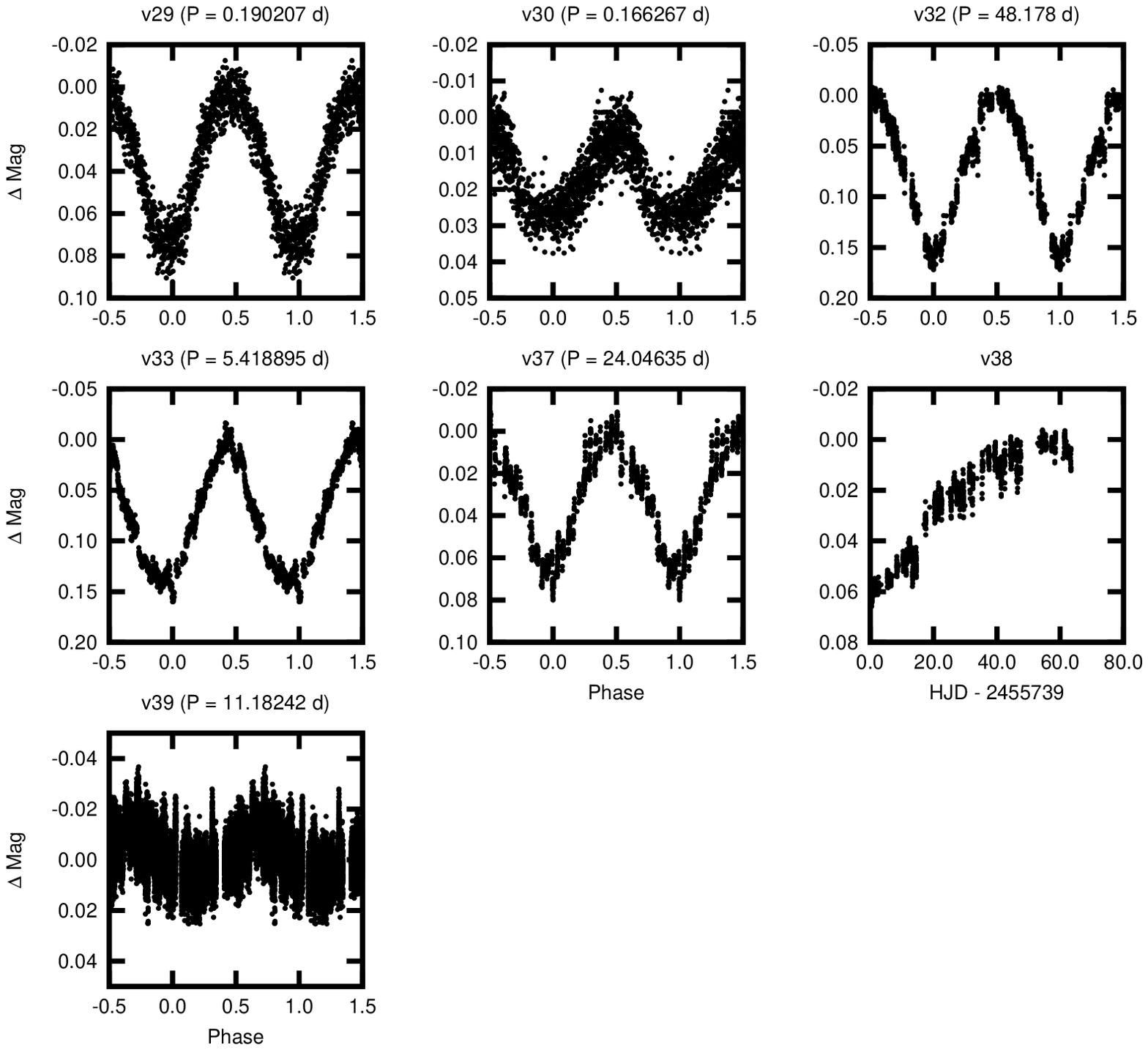}
\caption{Folded light curves for the newly-discovered, singly periodic $\delta$ Scuti-like variable stars v29 and v30, and the variable stars v32, v33, v37, v38 and v39. No significant periodicity could be determined for the variable v38 as less than one full cycle has clearly been observed. v39 displays large scatter in the data points on a nightly basis, the long-term periodic variations are more clearly seen unbinned.}
\label{figure:variables2}
\end{figure*}

{\bf{v25}}: A possible long-period grazing eclipsing binary (see Fig. \ref{figure:variables1}). Using PDM we find a period of \mbox{$P = 11.66332\pm0.00056$ d}. v25 is located in a very crowded region towards the centre of M71 (see Fig. \ref{figure:finders}). This, combined with its relatively long period and shallow eclipse depth \mbox{($\sim30$ mmag)} may explain why it was previously undetected. A possible X-ray active binary candidate \mbox{s07} \citep{0004-637X-687-2-1019,2010A&A...513A..16H} is located \mbox{$1\farcs13$} from v25 and its position is encompassed in the photometry aperture. 

{\bf{v26}}: A short period, low-amplitude variable with a best-fitting period of \mbox{$P=0.41370\pm0.00005$ d}, found using an LS periodogram. This object appears to be a $\delta$ Scuti-type variable but as its metallicity has not been determined here we cannot definitively distinguish between SX Phe and $\delta$ Scuti variations. For now we name v26 and the following four variables (v27, v28, v29 and v30), $\delta$ Scuti-like variable stars. Fourier analysis of v26 returns $2$ significant (SNR$>4$) peaks in the power spectrum $f_{1}=2.41849$ and $f_{2}=2.00666$ cycles d$^{-1}$ (see Fig. \ref{figure:v26-v31}). 

{\bf{v27}}: A low amplitude, $\delta$ Scuti-like variable. We find a period of \mbox{$P=3.571429\pm0.00123$ d} using a LS periodogram. When folded on this period v27 appeared to vary irregularly. Nightly inspection of the data showed variations on a shorter timescale. v27 was folded on the next significant period, \mbox{$P=0.77927\pm0.00027$ d}, where it shows regular, low-amplitude $\delta$ Scuti-like variations. A further two possible X-ray active binary candidates \mbox{s27a} and \mbox{s27b} \citep{0004-637X-687-2-1019,2010A&A...513A..16H} are located \mbox{$0\farcs03$} and \mbox{$0\farcs72$} from v27, respectively. Given the small angular separation we believe v27 to be X-ray binary candidate \mbox{s27a}. Fourier analysis of v27 returned many peaks with slightly reduced significance ($3<SNR_{f_{n}}<4$) which could indicate aperiodic variability. Regardless, Fig. \ref{figure:v26-v31} shows the variations of v27 along with the best fitting model created using the three most significant peaks $f_{1}=0.28104$ and $f_{2}=0.14195$ cycles d$^{-1}$. Although $f_{1}$ and $f_{2}$ appear to be multiples of one another, phasing upon either period individually results in a poorer fitting model.

{\bf{v28}}: A $\delta$ Scuti-like low-amplitude variable. The strongest peak in the LS periodogram of v28 was found at \mbox{$P = 0.082959\pm0.000002$ d}. The $\delta$ Scuti-like variations are also visible on a nightly basis with this period. v28 is the shortest period $\delta$ Scuti-like variable discovered here, with a period within $-1.47<\mathrm{log}P<-0.90$ d suggesting it could possibly be an SX Phe-type variable \citep{2011AJ....142..110M}. Fourier analysis of v28 returned $3$ significant (SNR$>4$) peaks in the power spectrum $f_{1}=12.05347$, $f_{2}=3.00624$ and $f_{3}=11.61858$ cycles d$^{-1}$ (see Fig. \ref{figure:v26-v31}).

\begin{figure*}
\rotatebox{270}{\includegraphics[width=12cm, height=17cm,trim=0mm 0mm 0mm 00mm]{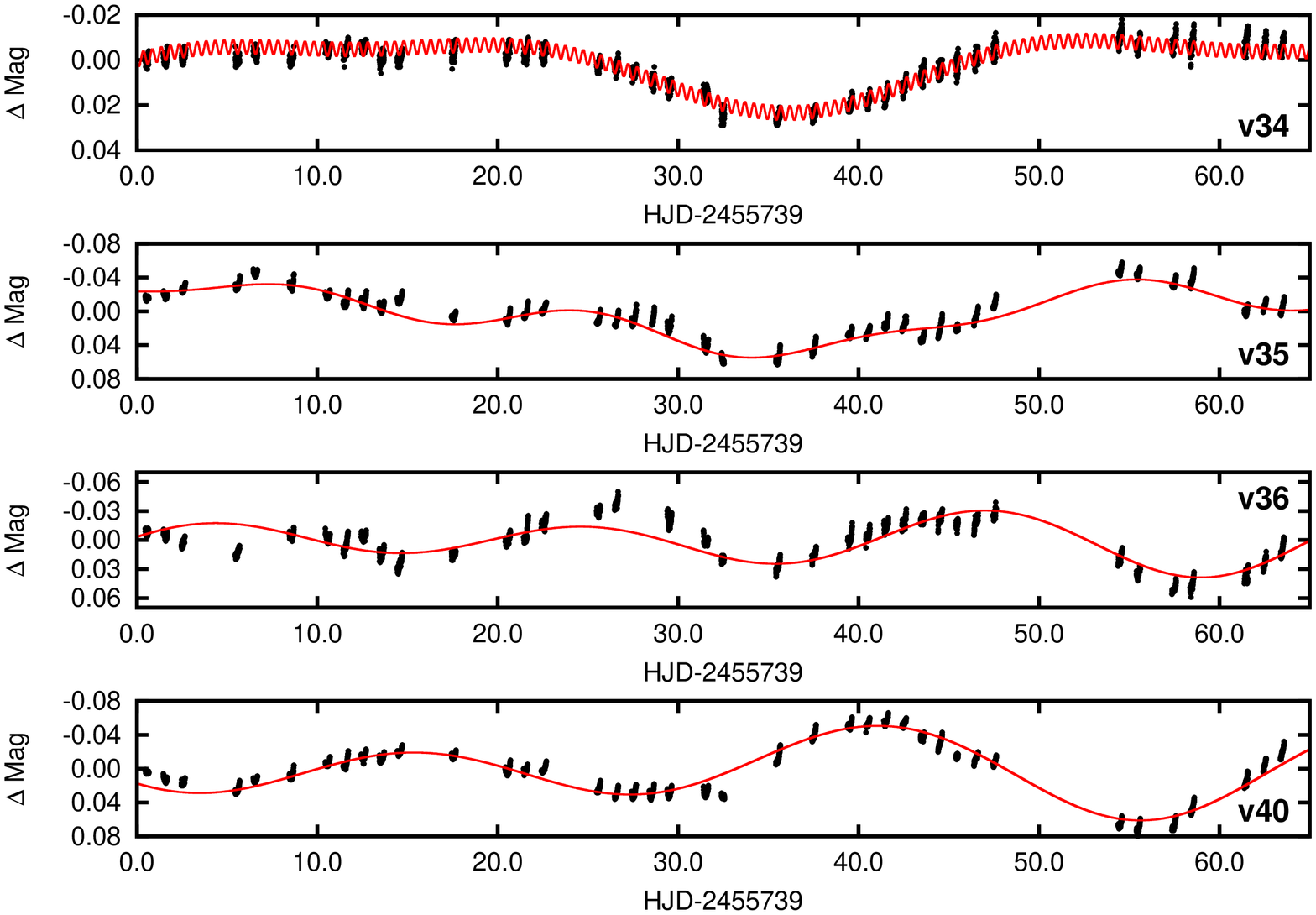}}
\caption{Light curves for the multiply periodic variables v34, v35, v36 and v40, over plotted with the best fitting models from the Fourier analyses. The data have been combined into $50$ binned points per night.The significant frequencies detected for each object are given in Table \ref{table:freqs}. }
\label{figure:v34-v40}
\end{figure*}

{\bf{v29}}: A larger amplitude \mbox{($\Delta V\sim0.07$ mag)} $\delta$ Scuti-like variable (see Fig. \ref{figure:variables2}). We find a best-fitting period of \mbox{$P = 0.190207\pm0.000017$ d} using PDM. Variations on this period are clearly visible in a night-by-night inspection of the data.

{\bf{v30}}: The final $\delta$ Scuti-like variable discovered here. We find a period of \mbox{$P=0.166267\pm0.000018$ d} using PDM. v30 was seen to vary only on the nights towards the end of the observing campaign and requires additional observations to confirm the type of variability (see Fig. \ref{figure:variables2}).

{\bf{v31}}: A possible eclipsing binary with asymmetrical eclipses. Using PDM we find a best-fitting period of \mbox{$P = 4.8940\pm0.0011$ d}.  Analysis of the Fourier power spectrum of v31 returned peaks at $f_{1}=0.40896$ and $f_{2}=0.20367$ cycles d$^{-1}$. Similarly to v27, $f_{1}$ and $f_{2}$ appear to be multiples of each other but both frequencies are required to fit the double peaked variation shown in Fig. \ref{figure:v26-v31}.

{\bf{v32}}: A relatively faint long-period variable star of undefined type (see Fig. \ref{figure:variables2}). We find a best fitting period of \mbox{$P = 48.1782\pm0.0206$ d} using PDM. It is possible that less than one full cycle of this star has been measured.

{\bf{v33}}: A long period variable star of undefined type (see Fig. \ref{figure:variables2}). We detect a significant period at \mbox{$P= 5.418895\pm0.00018$ d} using a LS periodogram.

{\bf{v34}}: A possible long period eclipsing binary. Analysis of v34 using a Fourier power spectrum returns $3$ significant frequencies $f_{1}=0.02182$, $f_{2}=0.04191$ and $f_{3}=2.00960$ cycles d $^{-1}$ (see Fig. \ref{figure:v34-v40}). v34 also displays short period variations, however the periods of which could not be detected with any significance. v34 appears to be the most complexly variable system discovered here. 

{\bf{v35}}: A long period, possible semi-regular variable. v35 is much brighter than the MS in M71 and is possibly a pulsating giant or a field star. Fourier analysis of v35 returns several significant (SNR$>4$) peaks in the power spectrum $f_{1}=0.01822$, $f_{2}=0.06496$ and $f_{3}=0.03961$ cycles d$^{-1}$. The light curve of v35 appears to be best fitted by a combination of them all (see Fig. \ref{figure:v34-v40}).

{\bf{v36}}: A bright variable star of undefined type. Our LS analysis returned a period of \mbox{$P\sim$$20$ d}. A search for aliases around this period showed that not all the variations were of equal strength. The power spectrum of v36 shows $2$ significant peaks at $f_{1}=0.04437$ and $f_{2}=0.03407$ cycles d$^{-1}$. It is clear that the object is variable (see Fig. \ref{figure:v34-v40}) but the nature of the variability is uncertain. Further observations are required to constrain any periodicity.

{\bf{v37}}: A long-period variable star of undefined type (see Fig. \ref{figure:variables2}). We find a best-fitting period of \mbox{$P=24.04635\pm0.00083$ d}. The X-ray binary \mbox{s20} \citep{0004-637X-687-2-1019,2010A&A...513A..16H} is located \mbox{$0\farcs07$} from v37 and encompassed by the photometry aperture.

\begin{table}
\caption{A summary of the significant frequencies detected in the multi-periodic light curves of $9$ stars observed during the NITES survey of M71. All except QU Sge are newly discovered here. }
\label{table:freqs}
\begin{tabular}{@{}lccc}
\hline\hline
Star & $f_{1}$ & $f_{2}$ & $f_{3}$  \\
 & cycles d$^{-1}$ & cycles d$^{-1}$ & cycles d$^{-1}$ \\
\hline
QU Sge	& 33.9167		& 36.1381 & 38.8864 	\\
v26		& 2.41849		& 2.00666 & ---			\\
v27		& 0.28104		& 0.14195 & ---			\\
v28		& 12.0535		& 3.00624 & 11.61858	\\
v31		& 0.40896		& 0.20367 & --- 		\\
v34		& 0.02182		& 0.04191 & 2.00960 	\\
v35		& 0.01822		& 0.06496 & 0.03961 	\\
v36		& 0.04437		& 0.03407 & --- 		\\
v40		& 0.03723		& 0.02534 & ---			\\ 
\end{tabular}
\end{table}

\begin{figure*}
\includegraphics[width=15cm, height=15cm]{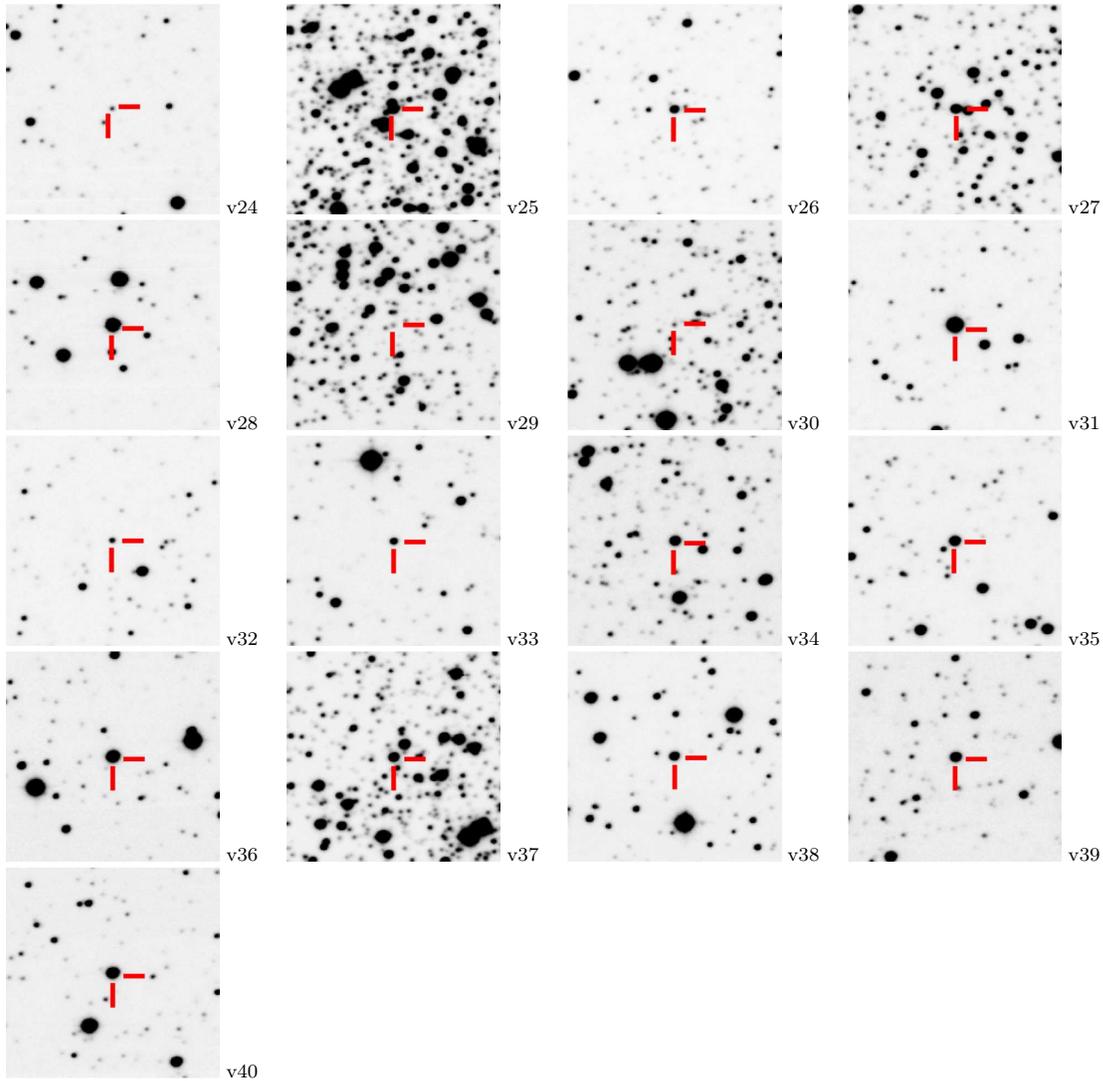}
\caption{Finding charts for the $17$ new variables (v24 - v40) discovered here. The charts are each $1\arcmin\times1\arcmin$ and taken from the WFC $V$ data described in \S\ref{subsec:WFCObservations}. North is up and East is left. }
\label{figure:finders}
\end{figure*}

{\bf{v38}}: A very long period variable, possibly the longest detected here (see Fig. \ref{figure:variables2}). We find one significant peak ($f_{1}=0.00532$ cycles d$^{-1}$, $P\approx188$ d) in the Fourier power spectrum of v38, however this is much longer than the length of the NITES survey hence constraining the period is not possible. Continued observation of v38 is required to constrain the period.

{\bf{v39}}: A long period low-amplitude variable of undefined type. A best-fitting period of \mbox{$P=11.18242\pm0.00051$ d} was found using a LS periodogram. The data shows a large intra-nightly scatter and binning masks the smooth long term variations seen unbinned (see Fig. \ref{figure:variables2}).

{\bf{v40}}: A possible semi-regular variable. Analysis of the power spectrum of v40 reveals $2$ significant peaks at $f_{1}=0.03723$ and $f_{2}=0.02534$ cycles d$^{-1}$ (see Fig. \ref{figure:v34-v40}). It appears from Fig. \ref{figure:v34-v40} that additional frequencies are present in v40 but are not detected with sufficient significance in our survey data.

\begin{table*}
\caption{Summary of the new and previously discovered variables observed in the NITES survey. The distance from the cluster centre r$_{\mathrm{c}}$ in arcseconds was calculated for each star using the cluster centre coordinates \mbox{R.A. (J2000) 19$^{\mathrm{h}}$53$^{\mathrm{m}}$46$^{\mathrm{s}}$} and \mbox{Dec. (J2000) +18$^{\mathrm{\circ}}$46\arcmin40\arcsec} of PN00. \mbox{$\Delta V$} is the peak-to-peak change in brightness in magnitudes and $N_{\mathrm{obs}}$ is the final number of reduced observations analysed for each target. The periods for multiply periodic variables are the principal periods detected. A period improvement is defined as yes if it has been newly determined here or significantly refined with respect to previously published values. (DB) detached binary; (SD) semi-detached binary; (CB) contact binary;  (EB) eclipsing binary of undefined type; ($\delta$ Sct-like) $\delta$ Scuti-like variable star; (LP) Long period variable of undefined type; (SR) Semi-regular type variable; (MP) multiply periodic.}
\label{table:3}
\begin{tabular}{@{}lccccccccc}
\hline\hline
Star & R.A. & Dec. & r$_{\mathrm{c}}$ & $\Delta$V & N$_{\mathrm{obs}}$ & Period & P$_{\mathrm{err}}$ &  Period & Variable \\
 & (J2000) & (J2000) & (arcsec) & (mag) & & (d) & (d) & improved & type \\
\hline
\multicolumn{10}{l}{Previously discovered:} \\
\mbox{\emph{QU Sge}}	& 19$^{\mathrm{h}}$53$^{\mathrm{m}}$49$^{\mathrm{s}}$.34 & +18$^{\mathrm{\circ}}$45\arcmin43.26\arcsec & 74.0 & 0.892 & 23\,466 & 3.79100 & 0.00061 & No   & SD / MP \\ 
\emph{S4}	& 19$^{\mathrm{h}}$54$^{\mathrm{m}}$04$^{\mathrm{s}}$.91 & +18$^{\mathrm{\circ}}$47\arcmin24.55\arcsec & 272.2 & 0.367 & 22\,229 & 2.525252 & 0.000033 & Yes  & DB \\ 
\emph{v1}	& 19$^{\mathrm{h}}$53$^{\mathrm{m}}$57$^{\mathrm{s}}$.49 & +18$^{\mathrm{\circ}}$43\arcmin33.76\arcsec & 247.6 & 0.573 & 23\,243 & 0.348904 & 0.000025 & No   & CB \\ 
\emph{v2}	& 19$^{\mathrm{h}}$53$^{\mathrm{m}}$57$^{\mathrm{s}}$.09 & +18$^{\mathrm{\circ}}$45\arcmin46.97\arcsec & 166.2 & 0.357 & 23\,213 & 0.36721   & 0.00001   & No   & CB \\ 
\emph{v3}	& 19$^{\mathrm{h}}$53$^{\mathrm{m}}$50$^{\mathrm{s}}$.84 & +18$^{\mathrm{\circ}}$47\arcmin51.48\arcsec & 99.2   & 0.168 & 23\,238 & 0.373856 & 0.000018 & No   & CB \\ 
\emph{v4}	& 19$^{\mathrm{h}}$53$^{\mathrm{m}}$49$^{\mathrm{s}}$.34 & +18$^{\mathrm{\circ}}$47\arcmin49.70\arcsec & 84.3   & 0.096 & 23\,206 & 0.55615   & 0.00002   & No   & DB  \\ 
\emph{v5}	& 19$^{\mathrm{h}}$53$^{\mathrm{m}}$34$^{\mathrm{s}}$.28 & +18$^{\mathrm{\circ}}$44\arcmin05.01\arcsec & 227.4 & 0.373 & 23\,240 & 0.404341 & 0.000015 & No   & CB  \\ 
\emph{v10}	& 19$^{\mathrm{h}}$54$^{\mathrm{m}}$01$^{\mathrm{s}}$.76 & +18$^{\mathrm{\circ}}$47\arcmin16.73\arcsec & 226.8 & 0.106 & 21\,642 & 0.796674 & 0.000033 & Yes  & CB? \\ 
\emph{v16}	& 19$^{\mathrm{h}}$53$^{\mathrm{m}}$58$^{\mathrm{s}}$.96 & +18$^{\mathrm{\circ}}$49\arcmin28.48\arcsec & 249.5 & 0.062 & 23\,227 & 0.642204 & 0.000047 & Yes  & CB? \\ 
\emph{v19}	& 19$^{\mathrm{h}}$54$^{\mathrm{m}}$09$^{\mathrm{s}}$.60 & +18$^{\mathrm{\circ}}$47\arcmin21.78\arcsec & 337.7 & 0.594 & 23\,174 & 6.1070     & 0.0042     & Yes  & EB? \\ 
\emph{v20}	& 19$^{\mathrm{h}}$53$^{\mathrm{m}}$23$^{\mathrm{s}}$.44 & +18$^{\mathrm{\circ}}$42\arcmin10.61\arcsec & 418.6 & 0.540 & 9\,585   & 0.469967 & 0.000028 & Yes  & CB \\ 
\emph{v21}	& 19$^{\mathrm{h}}$53$^{\mathrm{m}}$25$^{\mathrm{s}}$.54 & +18$^{\mathrm{\circ}}$51\arcmin17.44\arcsec & 401.7 & 0.330 & 23\,129 & 0.357907 & 0.000005 & Yes  & CB \\ 
\emph{v23}	& 19$^{\mathrm{h}}$53$^{\mathrm{m}}$57$^{\mathrm{s}}$.68 & +18$^{\mathrm{\circ}}$42\arcmin50.67\arcsec & 283.0 & 0.602 & 23\,293 & 5.657241 & 0.000027 & Yes  & DB \\ 
\multicolumn{10}{l}{Newly discovered:}  \\
\emph{v24}  & 19$^{\mathrm{h}}$54$^{\mathrm{m}}$08$^{\mathrm{s}}$.11 & +18$^{\mathrm{\circ}}$42\arcmin13.41\arcsec & 412.0  & 0.936 & 23\,268 & 1.33673     & 0.00029   & Yes & DB \\ 
\emph{v25}  & 19$^{\mathrm{h}}$53$^{\mathrm{m}}$46$^{\mathrm{s}}$.43 & +18$^{\mathrm{\circ}}$46\arcmin46.01\arcsec & 8.6      & 0.033 & 22\,690 & 11.66332   & 0.00056   & Yes & EB? \\ 
\emph{v26}  & 19$^{\mathrm{h}}$53$^{\mathrm{m}}$47$^{\mathrm{s}}$.94 & +18$^{\mathrm{\circ}}$49\arcmin37.19\arcsec & 179.3  & 0.045 & 23\,223 & 0.413481   & 0.000052   & Yes & $\delta$ Sct-like / MP \\ 
\emph{v27}  & 19$^{\mathrm{h}}$53$^{\mathrm{m}}$51$^{\mathrm{s}}$.48 & +18$^{\mathrm{\circ}}$46\arcmin00.53\arcsec & 87.3    & 0.020 & 23\,466 & 3.5582     & 0.0012   & Yes & $\delta$ Sct-like / MP \\ 
\emph{v28}  & 19$^{\mathrm{h}}$53$^{\mathrm{m}}$28$^{\mathrm{s}}$.40 & +18$^{\mathrm{\circ}}$41\arcmin45.31\arcsec & 386.5  & 0.013 & 23\,137 & 0.082964   & 0.000002 & Yes & $\delta$ Sct-like / MP \\ 
\emph{v29}  & 19$^{\mathrm{h}}$53$^{\mathrm{m}}$42$^{\mathrm{s}}$.72 & +18$^{\mathrm{\circ}}$45\arcmin58.35\arcsec & 62.5    & 0.070 & 23\,245 & 0.190207   & 0.000017 &Yes & $\delta$ Sct-like \\ 
\emph{v30}  & 19$^{\mathrm{h}}$53$^{\mathrm{m}}$53$^{\mathrm{s}}$.08 & +18$^{\mathrm{\circ}}$47\arcmin14.36\arcsec & 106.3  & 0.030 & 23\,190 & 0.166267   & 0.000018 & Yes & $\delta$ Sct-like \\ 
\emph{v31}  & 19$^{\mathrm{h}}$54$^{\mathrm{m}}$02$^{\mathrm{s}}$.86 & +18$^{\mathrm{\circ}}$42\arcmin50.31\arcsec & 331.8  & 0.112 & 17\,153 & 2.4452       & 0.0011      & Yes & EB? / MP     \\ 
\emph{v32}  & 19$^{\mathrm{h}}$54$^{\mathrm{m}}$07$^{\mathrm{s}}$.41 & +18$^{\mathrm{\circ}}$43\arcmin04.73\arcsec & 372.6  & 0.174 & 23\,355 & 48.178     & 0.021      & Yes & LP \\ 
\emph{v33}  & 19$^{\mathrm{h}}$53$^{\mathrm{m}}$28$^{\mathrm{s}}$.62 & +18$^{\mathrm{\circ}}$44\arcmin05.73\arcsec & 291.1  & 0.181 & 23\,240 & 5.418895   & 0.00018    & Yes & LP         \\ 
\emph{v34}  & 19$^{\mathrm{h}}$53$^{\mathrm{m}}$41$^{\mathrm{s}}$.39 & +18$^{\mathrm{\circ}}$44\arcmin44.41\arcsec & 132.8  & 0.043 & 22\,962 & 45.8295     & 0.0056      & Yes  & MP \\ 
\emph{v35}  & 19$^{\mathrm{h}}$53$^{\mathrm{m}}$45$^{\mathrm{s}}$.09 & +18$^{\mathrm{\circ}}$43\arcmin01.35\arcsec & 219.0  & 0.114 & 22\,486 & 54.88470     & 0.00056       & Yes & SR? / MP \\ 
\emph{v36}  & 19$^{\mathrm{h}}$54$^{\mathrm{m}}$11$^{\mathrm{s}}$.28 & +18$^{\mathrm{\circ}}$52\arcmin01.56\arcsec & 481.9  & 0.098 & 19\,828 & 22.5377     & 0.0016       & Yes  & SR? / MP \\ 
\emph{v37}  & 19$^{\mathrm{h}}$53$^{\mathrm{m}}$48$^{\mathrm{s}}$.95 & +18$^{\mathrm{\circ}}$47\arcmin13.76\arcsec & 53.8    & 0.080 &  20\,905 & 24.04635  & 0.00083    & Yes & LP \\ 
\emph{v38}  & 19$^{\mathrm{h}}$54$^{\mathrm{m}}$05$^{\mathrm{s}}$.44 & +18$^{\mathrm{\circ}}$51\arcmin58.36\arcsec & 421.3  & 0.067 & 22\,695 & ---	      & ---             & ---  & LP  \\ 
\emph{v39}  & 19$^{\mathrm{h}}$53$^{\mathrm{m}}$38$^{\mathrm{s}}$.34 & +18$^{\mathrm{\circ}}$49\arcmin25.33\arcsec & 197.9  & 0.021 & 22\,732 & 11.18242   & 0.00051   & Yes & LP   \\ 
\emph{v40}  & 19$^{\mathrm{h}}$53$^{\mathrm{m}}$41$^{\mathrm{s}}$.90 & +18$^{\mathrm{\circ}}$51\arcmin24.81\arcsec & 290.7  & 0.147 & 20\,932 & 26.86000   & 0.00062   & Yes  & SR? / MP \\  
\end{tabular}
\end{table*}

\subsection{No RR Lyrae stars in M71?}
\label{subsec:LackOfRRLyr}

One striking thing to note about M71 is the lack of RR Lyrae (RRL) variable stars discovered in previous surveys. RRL variables are old, metal-poor, population II, pulsating horizontal branch stars and make up a significant fraction of all the variables known in GCs. M71 is an intermediate population cluster and is relatively metal rich (only M69 has a higher metallicity), which likely contributes to the lack of RRL stars. The colour-magnitude diagram of AH71 places the Horizontal Branch (HB) of M71 at \mbox{V$\approx14.5$} which is well above the sensitivity limit of the NITES survey. However, the HB is relatively short and quite red so that a potential RRL progenitor in M71 may never evolve blue enough on the HB to reach the RRL instability strip. 

The QUEST RRL survey \citep{2004AJ....127.1158V} found that fundamental mode (RRLab) and first overtone (RRLc) variables have mean $V$ band amplitudes and periods of $1.04\pm0.24$ and $0.536\pm0.13$ mags and $0.539\pm0.009$ and $0.335\pm0.007$ d, respectively. The unbinned data in Fig. \ref{figure:rmsvsmag} (black points) shows that the NITES telescope has a typical photometric accuracy of $\sim5$ mmag at HB magnitudes in M71. The faintest stars surveyed here ($V\approx19$ mag) typically have an $RMS\leq0.1$ mag. Therefore, given that the amplitude of variation of RRLab and RRLc stars is several orders of magnitude larger than our photometric performance on the HB of M71, and the high time resolution of our survey data, we conclude that the cluster is devoid of RRL stars with amplitudes $\Delta V\geq 0.1$ mag, within the field surveyed ($11\arcmin\times11\arcmin \approx 0.6r_{t}$). Further observation of the outer regions ($0.6r_{t}\leq r \leq1r_{t})$ of M71 are required to determine if M71 is completely devoid of RRL variable stars.

\section{ESTIMATING SPECTRAL TYPES AND MEMBERSHIP PROBABILITIES OF VARIABLE STARS TOWARDS M71}
\label{sec:SpectralTypes}

\begin{figure}
\includegraphics[width=8cm, height=12cm, trim= 10mm 0mm 15mm 0mm]{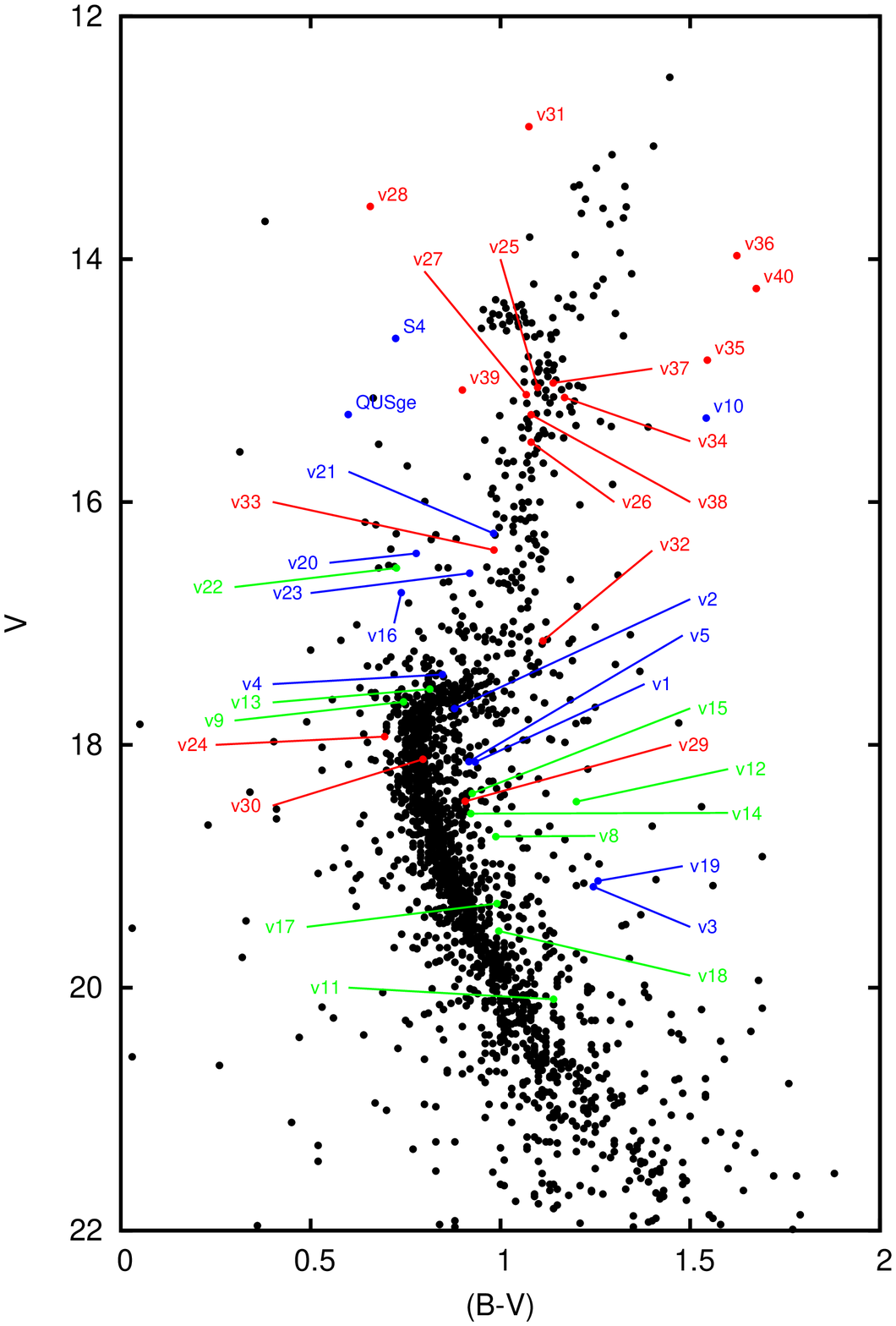}
\caption{Figure 5 of GM00 over plotted with the photometry of the variable star population measured from our WFC data. Stars in red are newly discovered variable stars, those in blue are the previously known objects and those in green are previously known objects that were not observed in the NITES survey.}
\label{figure:M71HR}
\end{figure}

In order to constrain the spectral types of the variables observed in our survey we obtained $BV$ photometry of M71 using the WFC on the INT. GM00 present $BV$ photometry of the cluster but do not cover all of the variable stars described in \S\ref{subsec:StellarVariablesPrevious} and \S\ref{subsec:StellarVariablesNew}. We plot the position of each variable star, as observed in our $BV$ data in Fig. \ref{figure:M71HR}. Stars brighter than $V=17$ in fig. 5 of GM00 were probable cluster members as determined by their proper motion analysis. Stars fainter than $V=17$ were taken from \citet{1992AJ....103..460H}. 

\citet{1984Ap&SS.100...95J} measure variations in reddening towards M71 of $0.12< E(B-V)<0.32$. The Galactic reddening maps derived from the IRAS and COBE infrared sky surveys \citep{1998ApJ...500..525S}\footnotemark\footnotetext{http://irsa.ipac.caltech.edu/applications/DUST/} also show a variation in reddening across the face of the cluster. We assume the reddening towards each variable star of \citet{1998ApJ...500..525S} as the average reddening in a $5\arcmin\times5\arcmin$ area surrounding each object, see Table \ref{table:4}. The average reddening from $14$ previous surveys of M71 in several photometric systems, $10$ from \citet{1984Ap&SS.100...95J} and references within, and those of GM00, AH71, \citet[hereafter CW85]{1985AJ.....90...65C} and \citet{2002A&A...395..481G} gives $E(B-V)=0.29$. The \citet{1998ApJ...500..525S} maps indicate a mean reddening of $E(B-V)=0.305$ towards M71 \citep{2002A&A...395..481G}, which is in agreement with the general consensus of previous studies. We calculate the intrinsic $(B-V)_{0}$ and $V_{0}$ for each variable star using the values of \citet{1998ApJ...500..525S}, assuming $A_{V}/E(B-V) = 3.1$ and subsequently estimate a small range of spectral types for each variable using the transformations of \citet{Schmidt-Kaler1982}, see Table \ref{table:4}. As the spectral type conversions of \citet{Schmidt-Kaler1982} require knowledge of the star's luminosity class we quote a range of spectral types, for which the main-sequence and giant star transformations of \citet{Schmidt-Kaler1982} encompass the intrinsic $(B-V)_{0}$ colour. Spectroscopy of the variable stars towards M71 is needed to confirm their spectral types, distances and cluster membership probabilities more rigorously. 

v36 appears on Simbad as an M2.5 giant  (the only variable star in M71 with a known spectral type) but with no reference to its classification. CW85 determines S4 (V4) to be a field star based on its proper motion and he calculates that QU Sge (2-255, V3) has a marginal cluster membership probability of 51\%. \citet{1981AJ.....86..669L} also show QU Sge to be a nonmember based on its radial velocity but \citet{2006ApJ...636L.129J} argue against this. In \S\ref{subsec:StellarVariablesPrevious} we also determine QU Sge to be a non-member based on the period-luminosity relation of the SX Phe pulsating component of the system. Of the remaining variables presented here only $3$ others were included in the CW85 proper motion survey of M71. v22 (KC 2-216), v34 (KC 1-91) and v37 (KC 1-38) have cluster membership probabilities of $15$, $0$ and $79$\%, respectively. The additional names in parentheses given above for S4, QU Sge, v22, v34 and v37 are those of CW85.

\begin{table*}
\caption{Variable star parameters summary. Intrinsic $V_{0}$ magnitudes and $(B-V)_{0}$ colours for $38$ of the $40$ variable stars towards M71 calculated using the reddening measurements of \citet{1998ApJ...500..525S} and $A_{V}/E(B-V) = 3.1$. Stars v6 and v7 could not be identified with confidence in our WFC data. The range of spectral types have been estimated using the transformations of \citet{Schmidt-Kaler1982}. We note that the specific spectral type of each object depends upon the determination of its luminosity class, which in turn requires a full spectroscopic classification not carried out here. In the notes column the Distances ($d$ in kpc) to the short period $\delta$ Scuti-type variables are calculated using $V_{0}$ from Column 1 and Eq 1 of \citet{2011AJ....142..110M}, and the cluster membership probabilities ($p$ \%) are those from the proper motion survey of \citet{1985AJ.....90...65C}.}
\label{table:4} 
\begin{tabular}{@{}lccccc}
\hline\hline
Star & V$_{\mathrm{0}}$ & (B-V)$_{\mathrm{0}}$ & E(B-V) & Spectral  & Notes \\
 & (mag) & (mag) & (mag)  & Type  &   \\
\hline
\multicolumn{6}{l}{Previously discovered:} \\
QU Sge   & $14.321\pm0.018$ & $0.290\pm0.019$	 & $0.309\pm0.015$   &	F0 		& $d=1.68\pm0.07$, p=$51\%$ \\
S4	& $13.704\pm0.013$	 & $0.418\pm0.014$	 & $0.306\pm0.011$   &		F4/F5  	& $p=0\%$	 \\	
v1	& $17.212\pm0.017$	 & $0.634\pm0.020$	 & $0.300\pm0.012$   & 		G0/G2 	&   \\	
v2	& $16.755\pm0.017$	 & $0.573\pm0.160$	 & $0.306\pm0.012$   & 		G0/F9  	&   \\	
v3	& $18.190\pm0.035$	 & $0.930\pm0.057$	 & $0.316\pm0.016$   & 		G8/K2  	&  \\
v4	& $16.449\pm0.021$	 & $0.534\pm0.024$	 & $0.313\pm0.016$   & 		F8  		&   \\
v5	& $17.119\pm0.022$	 & $0.588\pm0.025$	 & $0.329\pm0.019$   & 		F9/G0  	&   \\
v8	& $17.830\pm0.026$	 & $0.689\pm0.031$	 & $0.299\pm0.014$   &		 G1/G5  	&   \\
v9	& $16.683\pm0.027$	 & $0.434\pm0.031$	 & $0.311\pm0.015$   & 		F5  		&   \\
v10	& $14.359\pm0.014$	 & $1.236\pm0.014$	 & $0.306\pm0.012$   &		 K3/K6  	&   \\
v11	& $19.129\pm0.045$	 & $0.828\pm0.079$	 & $0.312\pm0.012$   &		 G4/K0  	&   \\
v12	& $17.401\pm0.032$	 & $0.856\pm0.037$	 & $0.344\pm0.023$   &	 	G5/K1  	&   \\
v13	& $16.210\pm0.029$	 & $0.385\pm0.031$	 & $0.429\pm0.027$   &	 	F3  		&   \\
v14	& $17.511\pm0.029$	 & $0.582\pm0.033$	 & $0.340\pm0.024$   &	 	F9/G0  	&   \\
v15	& $17.359\pm0.028$	 & $0.590\pm0.031$	 & $0.336\pm0.022$   &	 	F9/G0  	&   \\
v16	& $15.744\pm0.016$	 & $0.416\pm0.017$	 & $0.324\pm0.014$   &	 	F4  		&   \\
v17	& $18.277\pm0.045$	 & $0.659\pm0.061$	 & $0.332\pm0.016$   &		 G0/G4  	&   \\
v18	& $18.630\pm0.031$	 & $0.705\pm0.045$	 & $0.291\pm0.014$   &		 G1/G6  	&   \\
v19	& $18.184\pm0.026$	 & $0.955\pm0.042$	 & $0.302\pm0.012$   &	 	G9/K3  	&   \\
v20	& $15.419\pm0.022$	 & $0.454\pm0.023$	 & $0.324\pm0.021$   &		 F6  		&   \\
v21	& $15.028\pm0.034$	 & $0.585\pm0.036$	 & $0.397\pm0.030$   &	 	F9/G0  	&   \\
v22	& $15.586\pm0.016$	 & $0.417\pm0.019$	 & $0.309\pm0.015$   &	 	F4  		& $p=15\%$ \\
v23	& $15.666\pm0.015$	 & $0.622\pm0.016$	 & $0.297\pm0.012$   &	 	F9/G2  	&   \\
\multicolumn{6}{l}{Newly discovered:}  \\
v24	& $17.054\pm0.019$	 & $0.411\pm0.021$	 & $0.284\pm0.015$   &		F4  		& \\
v25	& $14.044\pm0.019$	 & $0.772\pm0.019$	 & $0.326\pm0.018$   &	 	G2/G9  	& \\
v26	& $14.461\pm0.022$	 & $0.743\pm0.023$	 & $0.338\pm0.021$   &	 	G2/G8  	& $d=8.56^{+0.29}_{-0.28}$  \\
v27	& $14.159\pm0.016$	 & $0.759\pm0.018$	 & $0.309\pm0.014$   &	 	G2/G9  	& $d=10.75^{+0.36}_{-0.35}$  \\
v28	& $12.550\pm0.020$	 & $0.329\pm0.021$	 & $0.328\pm0.020$   & 		F1  		& $d=1.40^{+0.05}_{-0.05}$	 \\
v29	& $17.502\pm0.032$	 & $0.596\pm0.037$	 & $0.311\pm0.018$   &	 	F9/G1  	& $d=22.17^{+0.80}_{-0.77}$   \\
v30	& $17.146\pm0.022$	 & $0.483\pm0.026$	 & $0.313\pm0.015$   &	 	F6/F7  	& $d=17.41^{+0.60}_{-0.58}$  \\
v31	& $11.999\pm0.015$	 & $0.782\pm0.016$	 & $0.293\pm0.014$   &	 	G2/G9  	&  \\
v32	& $16.234\pm0.016$	 & $0.819\pm0.018$	 & $0.293\pm0.014$   &	 	G4/K0  	&  \\
v33	& $15.342\pm0.021$	 & $0.643\pm0.022$	 & $0.340\pm0.021$   &	 	G0/G3  	&  \\
v34	& $14.160\pm0.016$	 & $0.853\pm0.017$	 & $0.316\pm0.016$   &	 	G5/K1  	& $p=0\%$  \\
v35	& $13.891\pm0.015$	 & $1.242\pm0.015$	 & $0.303\pm0.014$   &	 	K3/K6  	&  \\
v36	& $12.966\pm0.021$	 & $1.299\pm0.021$	 & $0.324\pm0.013$   &	 	K3/K7  	&  \\
v37	& $14.008\pm0.018$	 & $0.813\pm0.018$	 & $0.326\pm0.017$   &	 	G4/K0   	& $p=79\%$ \\
v38	& $14.287\pm0.020$	 & $0.760\pm0.021$	 & $0.321\pm0.013$   &	 	G2/G9  	&  \\
v39	& $14.005\pm0.026$	 & $0.554\pm0.027$	 & $0.346\pm0.026$   &	 	F8/F9  	&  \\
v40	& $13.126\pm0.031$	 & $1.314\pm0.031$	 & $0.360\pm0.029$   &	 	K4/K7  	&  \\
\end{tabular}
\end{table*}

Using the period-luminosity relation of \citet{2011AJ....142..110M} (Eq 1) and $V_{0}$ we calculate distances in parsecs to each of the $\delta$ Scuti-like variables towards M71. Assuming the fundamental frequency has been detected we find that all but $1$ (v28) of the $\delta$ Scuti-like variables v26-v30 are located at large distances behind the cluster, see Table \ref{table:4}. v28 is the shortest period variable detected here, consistent with an SX Phe-type variability and is found to be in the foreground of the cluster at $d=1.40\pm0.05$ kpc. If the period which we have identified is the first over-tone rather than the fundamental, then the distances we have determined will be shorter by \mbox{$\sim$15\%}.

Within the area surveyed we find no evidence for RRL variables towards M71 hence the ratios of different types of variables in M71 differs significantly from the variable star ratios in GCs of \citet{2001AJ....122.2587C}. From the literature they estimate $\sim60$\% of variable stars in GCs to be RRL, $\sim4$\% are eclipsing binaries, $4$\% are SX Phe variables, $2$\% are Cepheids and $4$\% are semi-regular variables. Based on the current population of $40$ variable stars towards M71 we find that  $40$\% are eclipsing binaries, $5$\% are $\delta$ Sct-like, $2.5$\% semi-regular and $8$\% are long or multiply periodic variables of undefined type.

\section{SUMMARY}
\label{sec:Summary}

Our high-cadence observations of M71 have revealed an additional $17$ previously undiscovered variable stars towards the cluster, one of which is a detached eclipsing binary, $2$ are possible eclipsing binary systems, $5$ are $\delta$ Scuti-like variables, $5$ are long period variable stars of an as yet undefined type and $3$ are possible semi-regular variables. We confirm the variability of $13$ previously known variables and refine or calculate for the first time periods for $7$ of them. Our data shows that the non-member S4 is most likely a detached eclipsing binary and not an RRL variable as previously suggested, and v16 appears to show variations of differing depths, suggesting an eclipsing binary rather than an SX Phe-type variable as proposed by PN00. 

We estimate spectral types for $38$ of $40$ variables towards M71 (see Table \ref{table:4}), v6 and v7 which could not be identified with confidence in our WFC data and were subsequently excluded from our analysis. We cross correlate our catalogue of variable stars with the proper motion survey of CW85 and find cluster membership probabilities for $5$ variable stars (QU Sge, S4, v22, v34 and v37), $3$ of which are non-members and $2$ probable members. None have a high probability of belonging to M71. As M71 is located at low galactic latitude we expect increased contamination of variable stars from the field along the line of sight to the cluster, compared to other GCs in the halo. We estimate distances to $6$ variable stars towards M71 (QU Sge, v26, v27, v28, 29 and v30) using the period-luminosity relation of \citet{2011AJ....142..110M}, finding that all but one lie at large distances behind the cluster, assuming the fundamental mode of oscillation has been detected. Continued observation of M71 is also required to constrain any periodicity in $6$ of the newly discovered variable stars for which a period could not be constrained. 

Our search for transiting giant planets revealed $5$ significant transit-like detections. However none of these were consistent with a typical hot Jupiter signal. Two are the new short period, low amplitude variable stars v38 and v30, one is the probable eclipsing binary v10 of PN00 (which is confirmed here) and the light curves of the remaining two objects are inconsistent with giant transiting planet-like detections when folded on the periods detected. As only a small fraction of the MS in M71 was surveyed with the NITES telescope it is difficult to constrain the giant planet frequency in M71 with confidence. Continued observation of a larger fraction of the MS is required to place limits on the giant planet frequency. To increase the likelihood of detecting a transiting hot Jupiter in M71, or any GC for that matter, requires deeper observations of a more significant fraction of the cluster's main-sequence. In the case of M71, to do so would require photometry from a larger facility ($1-2$ m class telescopes, e.g. WFC on INT) to reach the photometric precision required to detect transiting giant planets at fainter magnitudes. 

We have shown that the NITES telescope is sensitive to low amplitude variability \mbox{($<0.02$ mag)} and well suited to its future role as a follow-up telescope for transiting exoplanets. The recent discovery of transiting Neptune-sized planets in the old OC NGC 6811 \citep{2013MeibomSNGC6811} has shown that planets do indeed survive in OCs. Hence, new surveys of younger, more metal-rich OCs using the NITES telescope could discover more transiting giant exoplanets in the future.

\section*{Acknowledgments}
This research has made use of the Simbad database, operated at CDS, Strasbourg, France. The authors would like to acknowledge the Isaac Newton Group of Telescopes for their support during the installation of the NITES telescope and Wojtek Pych for the use of his difference-imaging package DIAPL2.

\appendix 

\section{PHOTOMETRIC PERFORMANCE \& NOISE ANALYSIS OF THE NITES TELESCOPE}
\label{sec:PhotometricPerformance}

\begin{figure*}
\rotatebox{270}{\includegraphics[width=8cm, height=17cm,trim=30mm 10mm 40mm 10mm]{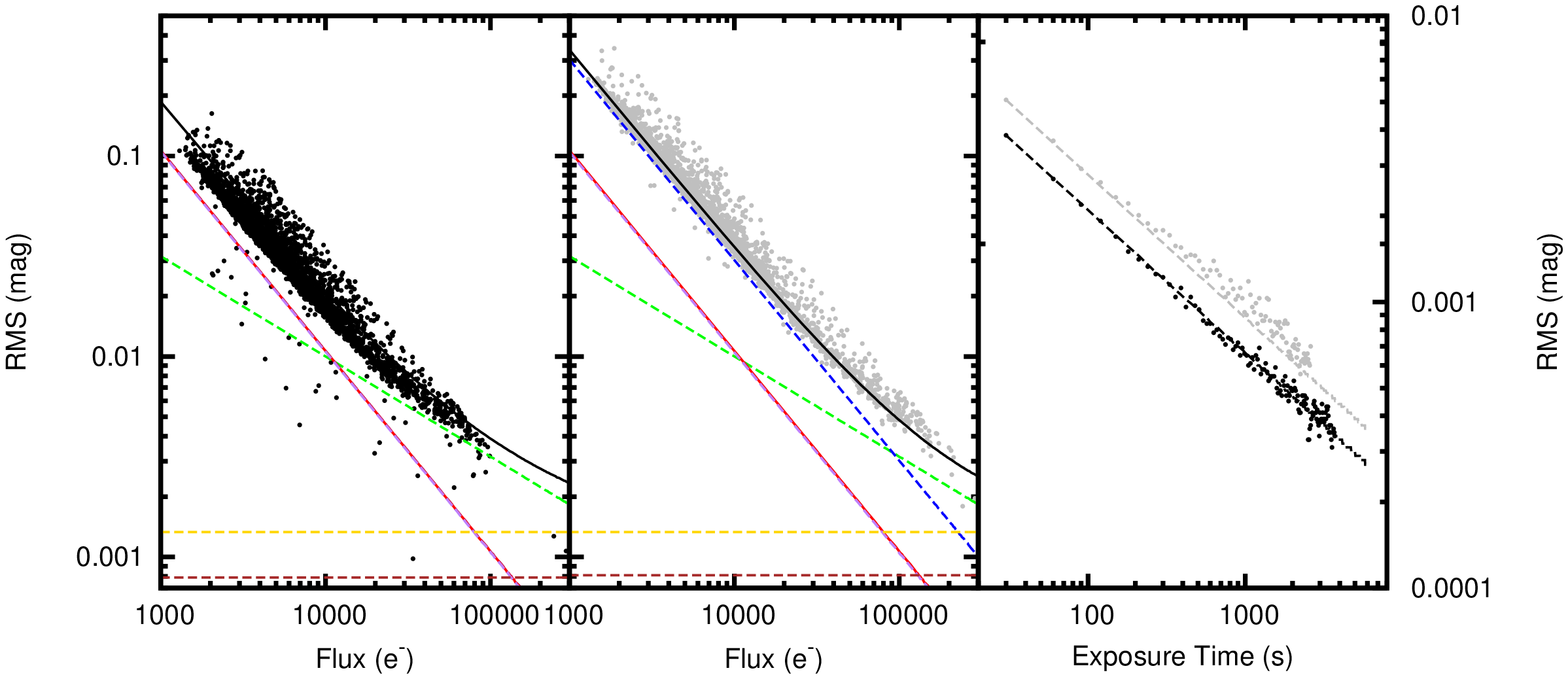}}
\caption{Black and grey points represent data taken during dark and bright time, respectively. The left and middle plots are RMS vs flux diagrams taken during dark and bright time, respectively. The right plot is the RMS vs binned exposure time for the average of the nine brightest non-variable stars during bright and dark time. The green, blue, red, purple, yellow and brown dashed lines represent the noise models from the target, sky, read noise, dark current, scintillation and flat fields, respectively. In the left panel the noise models of the sky, dark current and read noise (blue, purple and red dashed lines) are almost completely overlapping. In the central panel the noise models of the dark current and read noise (purple and red dashed lines) are also overlapping. The solid black line represents the total noise model, while the black and grey dashed lines show the expected $\sqrt{n}$ decrease in RMS in the presence of Gaussian noise only, where $n$ is the number of points per bin.}
\label{figure:noise}
\end{figure*}

During our survey of M71 the photometric performance of the system was measured using full moon and first quarter observations compared to a theoretical noise model. The noise model combines errors from the target, sky, read noise, dark current and scintillation with an estimation of the error in the flat fielding process using Eq \ref{eq:TotalNoiseModel};

\begin{equation}
N_{\mathrm{total}} = \sqrt{f + f_{\mathrm{sky}} + DC + \left(N_{\mathrm{r}}\right)^{2} + N_{\mathrm{flat}} + N_{\mathrm{sc}}^{2}}, \label{eq:TotalNoiseModel}
\end{equation}

\noindent where $f$, $f_{\mathrm{sky}}$, $DC$ and $N_{\mathrm{r}}$ are the flux from the target, sky, dark current and the read noise inside the photometry aperture, respectively and;

\begin{eqnarray}
N_{\mathrm{flat}}  &=& \frac{f \times n_{\mathrm{pix}}}{\sqrt{F_{\mathrm{total}}}},\\
N_{\mathrm{sc}} &=& 0.09D^{-2/3}\left(\sec\left(Z\right)\right)^{W}\mathrm{exp}\left(\frac{-h}{h_{\mathrm{0}}}\right)\left(2t\right)^{-1/2},
\end{eqnarray}

\noindent are the errors from flat fielding and the scintillation noise according to \citet{1998PASP..110..610D}, respectively. \mbox{$n_{\mathrm{pix}}$} is the number of pixels inside the photometric aperture, \mbox{$F_{\mathrm{total}}$} is the combined flux in \mbox{e$^{-}$ pix$^{-1}$} in the master flat field (typically \mbox{$\sim1\,000\,000$ e$^{-}$ pix$^{-1}$)}, $D$ is the diameter of the telescope aperture in cm, \mbox{$\sec\left(Z\right)$} is the airmass, $h$ is the altitude of the observatory in m, \mbox{$h_{\mathrm{0}} = 8\,000$ m} is the atmospheric scale height, $t$ is the integration time in s and $W$ is a variable dependent on angle between the line of sight and wind direction. \mbox{$W = 1.5$, $1.75$ or $2.0$} when observing perpendicular to the wind, close to the zenith and parallel to the wind, respectively.

The image subtracted frames (see \S\ref{subsec:NITESObservations}) typically have a background level with a mean flux \mbox{$\approx0$}, hence measuring the sky level directly is impossible. However, the noise from the sky will remain in the subtracted images and is expected to be Gaussian in nature. Hence, by measuring the standard deviation in the background of each subtracted image, the flux from the sky in the photometry aperture $f_{\mathrm{sky}}$ in e$^{-}$ can be estimated using:

\begin{equation}
f_{\mathrm{sky}}=\sigma_{\mathrm{sky}}^{2}\times G \times n_{\mathrm{pix}}, \label{eq:new_sky} 
\end{equation}

\noindent where $\sigma_{\mathrm{sky}}$ is the nightly average standard deviation of the background in the subtracted images and $G$ is the gain in $\mathrm{e}^{-}$ $\mathrm{ADU}^{-1}$.  

Figure \ref{figure:noise} (left and centre) show the noise model of the NITES telescope along with the typical photometric accuracy during dark and bright time, respectively. It is evident from Figure \ref{figure:noise} (right) that the noise in the system is essentially Gaussian as the data points follow the $\sqrt{n}$ decrease in RMS, where $n$ is the number of binned points. Figure \ref{figure:noise} (right) demonstrates that the system is relatively free from systematic noise down to the sub-mmag regime. A photometric accuracy of \mbox{$1$ mmag} is reached with binned exposure times of \mbox{$\approx400$ s} during dark time while bright time observations reach \mbox{RMS $<2$ mmag} on the same time scale. 

\label{lastpage}

\end{document}